\documentclass[journal]{IEEEtran}
\usepackage{cite}
\usepackage{amsmath}
\usepackage{amssymb}
\usepackage{algorithmic}
\usepackage{algorithm}
\usepackage{graphicx}
\usepackage{acronym}
\usepackage{textcomp}
\usepackage{xcolor}
\usepackage{colortbl}
\usepackage{multirow}
\usepackage{booktabs}
\usepackage{makecell}
\usepackage{orcidlink}

\usepackage{lipsum}
\usepackage{fancyhdr}
\fancypagestyle{sec}{\lhead{\tmpx}}

\fancypagestyle{arXiv}
{
   \fancyhf{}
   \chead{\textcolor{gray}{This paper has been accepted to be published in \\
   IEEE Journal on Emerging and Selected Topics in Circuits and Systems}}
   \cfoot{ \fontsize{=9pt}{10pt}\selectfont  \textcolor{gray}{© 2026 IEEE. Personal use of this material is permitted. Permission from IEEE must be obtained for all other uses, in any current or future media, including reprinting/republishing this material for advertising or promotional purposes, creating new collective works, for resale or redistribution to servers or lists, or reuse of any copyrighted component of this work in other works}\\\fontsize{=10pt}{12pt}\selectfont\thepage}
   
}

\providecommand{\BitFairTrackChanges}{false}
\providecommand{\BitFairShowChanges}{false}
\providecommand{\BitFairShowRev}{false}
\providecommand{\BitFairShowRevCG}{false}
\providecommand{\BitFairShowRevFinal}{true}

\newif\ifbitfairtrackchanges
\csname bitfairtrackchanges\BitFairTrackChanges\endcsname
\ifbitfairtrackchanges\else
  \renewcommand{\BitFairShowChanges}{false}
  \renewcommand{\BitFairShowRev}{false}
  \renewcommand{\BitFairShowRevCG}{false}
  \renewcommand{\BitFairShowRevFinal}{false}
\fi

\newif\ifshowchanges
\csname showchanges\BitFairShowChanges\endcsname
\ifshowchanges
  \DeclareRobustCommand{\changed}[1]{{\color{red}#1}}
  \DeclareRobustCommand{\newref}[1]{{\color{red}#1}}
\else
  \DeclareRobustCommand{\changed}[1]{#1}
  \DeclareRobustCommand{\newref}[1]{#1}
\fi
\newif\ifshowrev
\csname showrev\BitFairShowRev\endcsname
\ifshowrev
  \DeclareRobustCommand{\rev}[1]{{\color{blue}#1}}
\else
  \DeclareRobustCommand{\rev}[1]{#1}
\fi
\newif\ifshowrevcg
\csname showrevcg\BitFairShowRevCG\endcsname
\ifshowrevcg
  \DeclareRobustCommand{\revcg}[1]{{\color{blue}#1}}
\else
  \DeclareRobustCommand{\revcg}[1]{#1}
\fi
\newif\ifshowrevfinal
\csname showrevfinal\BitFairShowRevFinal\endcsname
\ifshowrevfinal
  \DeclareRobustCommand{\revfinal}[1]{{\color{blue}#1}}
\else
  \DeclareRobustCommand{\revfinal}[1]{#1}
\fi
\makeatletter
\AtBeginDocument{%
  \@ifpackageloaded{hyperref}{%
    \pdfstringdefDisableCommands{%
      \def\changed#1{#1}%
      \def\newref#1{#1}%
      \def\rev#1{#1}%
      \def\revcg#1{#1}%
      \def\revfinal#1{#1}%
    }%
  }{}%
}
\makeatother
\makeatletter
\g@addto@macro\fs@ruled{%
  \expandafter\def\expandafter\@fs@pre\expandafter{%
    \expandafter\begingroup\expandafter\normalcolor\@fs@pre}%
  \g@addto@macro\@fs@post{\endgroup}%
}
\makeatother

\newacro{XR}{Extended Reality}
\newacro{NN}{Neural Network}
\newacro{CNN}{Convolutional Neural Network}
\newacro{ReLU}{Rectified Linear Unit}
\newacro{BN}{Batch Normalization}
\newacro{MAC}{Multiply-Accumulate}
\newacro{MSB}{Most Significant Bit}
\newacro{LSB}{Least Significant Bit}
\newacro{ETR}{Early Termination Rate}
\newacro{PE}{Processing Element}
\newacro{ABO}{Adaptive Bit Ordering}
\newacro{SNN}{Spiking Neural Network}
\newacro{ANN}{Artificial Neural Network}
\newacro{VWW}{Visual Wake Words}
\newacro{FSM}{Finite State Machine}
\begin{document}

\title{\changed{BitFair: A 12-nm Bit-Serial CNN Accelerator with Learnable Early Termination and Adaptive Bit Ordering for Ultra-Low-Power XR Vision}}

\author{Ang~Li\orcidlink{0000-0003-3615-6755},
        Chang~Gao\orcidlink{0000-0002-3284-4078}%
\thanks{Corresponding author: Chang Gao (chang.gao@tudelft.nl).}%
\thanks{The authors are with the Department of Microelectronics, Delft University of Technology, Delft, The Netherlands. (Emails: ang.li@tudelft.nl, chang.gao@tudelft.nl)}
}

\maketitle

\begin{abstract}
\changed{\ac{XR} wearables require always-on perception within tight power envelopes of a few watts and motion-to-photon latency budgets below 20\,ms, leaving only a few milliseconds for neural-network inference.} Bit-serial computing is attractive for such energy-efficient neural network acceleration, but many existing architectures still process all bits even when ReLU sets the final output to zero. This paper presents BitFair, a software-hardware co-designed bit-serial \changed{CNN} accelerator with learnable bit-level early termination and adaptive bit ordering, working under \changed{the ultra-low-power and strict latency} requirements of XR applications. 
\revcg{BitFair exploits dynamic bit-level sparsity by learning per-layer thresholds that trigger early termination when partial sums reliably predict that the final ReLU output will be zero.}
Furthermore, it searches for layer-wise bit orders that prioritize informative bits, maximizing early termination without sacrificing accuracy. A 12-nm FinFET implementation with a core area of 0.34~mm$^2$, 104~KB on-chip memory, and voltage scaling from 0.55 to 0.70~V achieves sub-millisecond latency, up to \revfinal{234.0}~BTOPS/W, and 0.07~pJ/SOP. On IBM DVS128 Gesture and N-MNIST, BitFair achieves 96.5\% and 97.7\% accuracy, respectively, while improving effective energy efficiency by 4.0--22.1$\times$ and accuracy by up to 9.2\% over prior fabricated \changed{XR} vision accelerators.
\end{abstract}


\section{Introduction}
\thispagestyle{arXiv}
\label{sec:Intro}
In recent years, extended reality (XR), encompassing virtual reality (VR), augmented reality (AR), and mixed reality (MR), has driven a paradigm shift in computing, introducing natural interaction technologies such as gesture recognition and object detection~\cite{amir2017low-dvs-gesture, frenkel2022reckon, zhang202322}. To deliver immersive, interactive, and comfortable user experiences, XR wearables impose stringent constraints: sub-millisecond latency is required to support real-time responsiveness and prevent motion sickness, while ultra-low power consumption is critical for extending battery life in untethered devices~\cite{moin2021wearable, liu2019edge}. As XR applications increasingly rely on \acp{NN} for accurate perception, traditional algorithms and hardware encounter significant challenges in meeting these power and latency demands, highlighting the need for efficient hardware-software co-design~\cite{jacob2018quantization, lee2018unpu}.

Hand gesture recognition is a representative workload that crystallizes these constraints. Modern XR headsets such as Meta Quest~3 and Apple Vision Pro consume 5--30\,W at the system level~\cite{meta_quest3_power, apple_vp_specs}, leaving only tens of milliwatts for a dedicated neural-network accelerator that must operate continuously without draining the battery. The motion-to-photon latency budget, i.e., the total delay from user motion to the corresponding visual update, must remain below 20\,ms to prevent simulator sickness~\cite{mtp_latency}, of which only 2--5\,ms can be allocated to neural-network inference after accounting for sensor readout, rendering, and display scanout~\cite{liu2019edge}. \rev{Post-layout results using the GlobalFoundries 12-nm technology show} BitFair's inference latency of 0.12--1.55\,ms and power consumption of 0.8--13.7\,mW comfortably fit within both envelopes, making it a practical candidate for always-on XR perception. \rev{We use this term in a specific sense: continuous, on-device perception over sparse or low-resolution inputs under strict power and latency budgets. This is distinct from high-resolution, compute-heavy scene understanding. Our accelerator targets always-on lightweight visual perception workloads, such as digit recognition, hand gesture recognition, and person detection~\cite{chowdhery2019visual} using either event- or frame-based vision sensors.}

To address these demands, \acp{SNN} have emerged as a leading paradigm for energy-efficient, event-driven vision, as evidenced by several \changed{recently} fabricated accelerators targeting XR workloads~\cite{frenkel2022reckon, zhang202322, yang202540nm, fu2025neuc}. \acp{SNN} exploit temporal sparsity through integrate-and-fire dynamics: neurons remain silent in the absence of input spikes, naturally reducing switching activity and energy consumption. However, practical limitations constrain their broader deployment. Surrogate-gradient training complicates convergence in deeper architectures, and \acp{SNN} typically suffer an accuracy gap compared to equivalent \acp{CNN} on the same tasks. Furthermore, their efficiency advantage diminishes on dense inputs where spike rates approach saturation. These observations raise a fundamental question: can mature \acp{CNN}, with their well-understood training frameworks and outstanding baseline accuracy, achieve \ac{SNN}-like event-driven efficiency? 

\begin{figure}[t]
    \centering
    \includegraphics[width=0.99\linewidth]{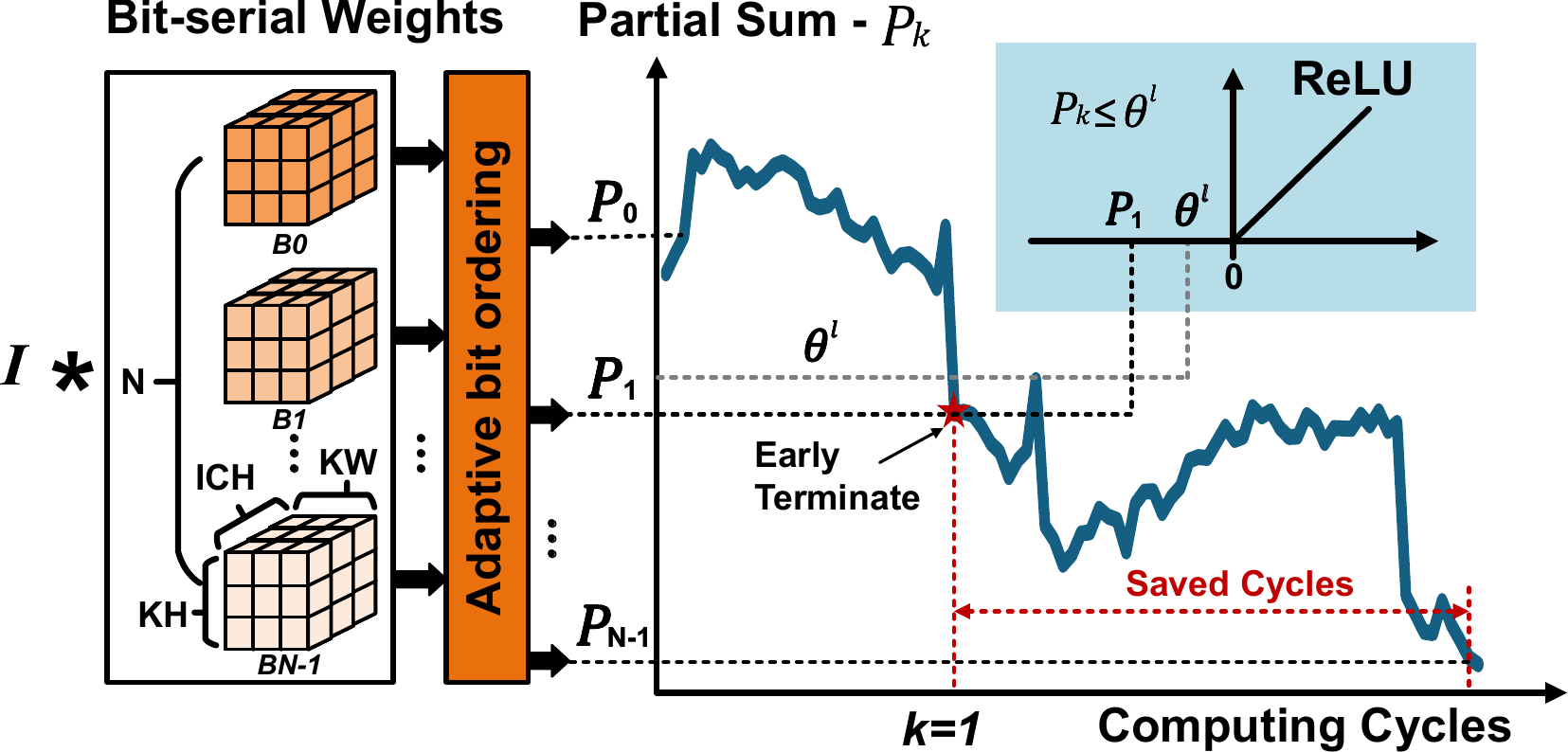}
    \caption{BitFair with early termination and adaptive bit ordering. Inputs are multiplied with bit-serial weights in an adaptive order to maximize early termination opportunities. 
    \revcg{When the partial sum $P_k$ falls below the layer-specific threshold $\theta^l$, the final ReLU output is predicted to be zero and computation terminates early, saving the processing of the remaining bits.}
    }
    \label{fig:idea}
\end{figure}

To deploy \acp{CNN} on resource-constrained edge devices, quantization techniques have been widely adopted to reduce computational costs by representing weights and activations with lower bit widths~\cite{jacob2018quantization, zhou2016dorefa}. However, conventional implementations process all bits atomically, leaving significant opportunities for bit-level optimization unexploited. Consequently, bit-serial accelerators~\cite{sharify2019laconic, albericio2017bit, judd2016stripes, lee2018unpu} have emerged as an energy-efficient alternative. Bit-serial architectures decompose operations into sequences of single-bit computations, enabling fine-grained control that can exploit bit-level sparsity within the binary representation of data to reduce \ac{MAC} energy consumption.

Despite these advances, existing bit-serial accelerators perform substantial redundant computation due to the structure of \acp{CNN}. ReLU activation functions clamp negative outputs to zero, inducing significant data-dependent runtime sparsity. However, most conventional bit-serial accelerators fail to exploit this dynamic sparsity, processing all non-zero bits regardless of the final result's sign. While previous works~\cite{lin2017predictivenet, song2018prediction, akhlaghi2018snapea} have proposed prediction-based early termination, these approaches operate at value-level granularity and are thus not directly compatible with bit-serial processing. BitSET~\cite{pan2023bitset} introduces an early termination mechanism for bit-serial architectures, but it is limited by the use of static thresholds and a rigid bit ordering.

This paper introduces \textbf{BitFair}, an \changed{ultra-low-power} bit-serial accelerator enhanced with early termination and adaptive bit ordering to satisfy the rigorous latency and power constraints of XR applications. Figure~\ref{fig:idea} illustrates the idea of BitFair. First, we propose learnable per-layer thresholds that optimize early termination decisions through gradient-based training, adapting to layer-specific characteristics. By employing soft thresholding with temperature annealing, we enable end-to-end differentiable optimization to maximize computational savings under strict accuracy constraints. Conceptually, this threshold-driven mechanism shares a similar spirit with the integrate-and-fire dynamics of \acp{SNN}: just as a spiking neuron remains silent and produces no output until its membrane potential crosses a firing threshold, BitFair's bit-serial computation terminates early and predicts a zero ReLU output whenever the evolving partial sum falls below the learned threshold.

Moreover, BitFair introduces an adaptive bit-ordering strategy that moves beyond the traditional \ac{MSB}-first approach. It performs a greedy search to determine a layer-specific bit processing order that enhances opportunities for early termination while prioritizing bits most critical to accuracy. For instance, when a layer's weights are predominantly clustered near $\pm 32$ (binary: 00100000), processing bit position 5 before bits 6 and 7 can trigger earlier termination (with bit positions indexed from LSB to MSB, where 0 denotes the LSB), resulting in computational savings without compromising accuracy.

The BitFair hardware accelerator is implemented in 12-nm FinFET technology to validate the proposed techniques. The design features a bit-serial \ac{PE} array with an output-stationary dataflow, effectively leveraging parallelism while accommodating the bit-level early termination mechanism. \rev{Post-layout evaluation shows} a compact core area of 0.34~mm$^2$, 104~KB of on-chip memory, and operation at a 500~MHz clock frequency, achieving sub-millisecond latency with a low power consumption of 13.7~mW.

The main contributions of this paper include:

\begin{enumerate}
\item \textbf{Bit-level early termination with learnable thresholds:} A trainable threshold mechanism that learns layer-wise early termination decisions through gradient-based optimization, replacing fixed-threshold methods to automatically balance efficiency and accuracy.
\item \textbf{Adaptive bit ordering optimization:} \changed{An} algorithm that adopts a greedy search strategy to determine a layer-wise bit processing order, further improving the early-termination rate and computational efficiency while preserving accuracy. 
\item \textbf{Hardware validation on XR-class deployment:} A hardware accelerator in GlobalFoundries 12-nm FinFET technology that achieves 0.34~mm$^2$ area and 13.7~mW power, compatible with wearable XR budgets, evaluated on both event-driven and frame-based vision classification benchmarks, and demonstrating 4.0$\times$--22.1$\times$ energy efficiency improvements and up to 9.2\% higher accuracy over prior fabricated \ac{SNN}-based edge vision accelerators.
\end{enumerate}

\section{Background and Motivation}
\label{sec:background}

\subsection{\changed{Spiking Neural Networks for XR Vision}}

The demand for ultra-efficient edge processing has made \acp{SNN} a highly attractive paradigm, driving \changed{significant} silicon investment in SNN-based accelerators~\cite{frenkel2022reckon, zhang202322, liu202430, yang202540nm, fu2025neuc}. Operating on \changed{an} integrate-and-fire model, neurons in these networks accumulate input spikes over discrete time steps and emit an output spike only when their membrane potential exceeds a firing threshold, after which the potential resets. The key appeal of this event-driven computation is its ability to inherently skip silent neurons, seamlessly achieving conditional computation without explicit prediction mechanisms.

Despite these advantages, \acp{SNN} face several practical limitations. The non-differentiable spike function necessitates surrogate gradient estimators during training~\cite{bengio2013estimating, neftci2019surrogate}, which complicate convergence for deeper networks and limit the model architectures that can be effectively trained~\cite{guo2023direct}. These training difficulties translate into a measurable accuracy gap: on DVSGesture, ReckOn~\cite{frenkel2022reckon} achieves 87.3\% and the recent CICC'25 design~\cite{yang202540nm} reaches 90.5\%, compared to 96.5\% for a CNN of comparable size (Table~\ref{tab:sota_comparison}). Furthermore, when processing dense RGB frames, high spike rates across all pixels saturate the temporal coding, degrading the efficiency advantage that \acp{SNN} enjoy on sparse event streams. \changed{A recent comprehensive analysis of digital hardware accelerators for both \acp{ANN} and SNNs~\newref{\cite{ottati2023spike}} corroborates these observations, demonstrating that on static vision tasks, \ac{ANN} accelerators Pareto-dominate their SNN counterparts in both energy efficiency and accuracy. While SNNs show competitive efficiency on temporal tasks, the study highlights that classification workloads employing bio-inspired sensors such as event cameras remain underexplored, motivating BitFair's approach of applying CNN-based processing to event-driven data.} These limitations have motivated exploring conditional computation within conventional CNN frameworks, where bit-serial processing offers a natural substrate for fine-grained computation control.

\begin{figure*}[t]
    \centering
    \includegraphics[width=0.94\linewidth]{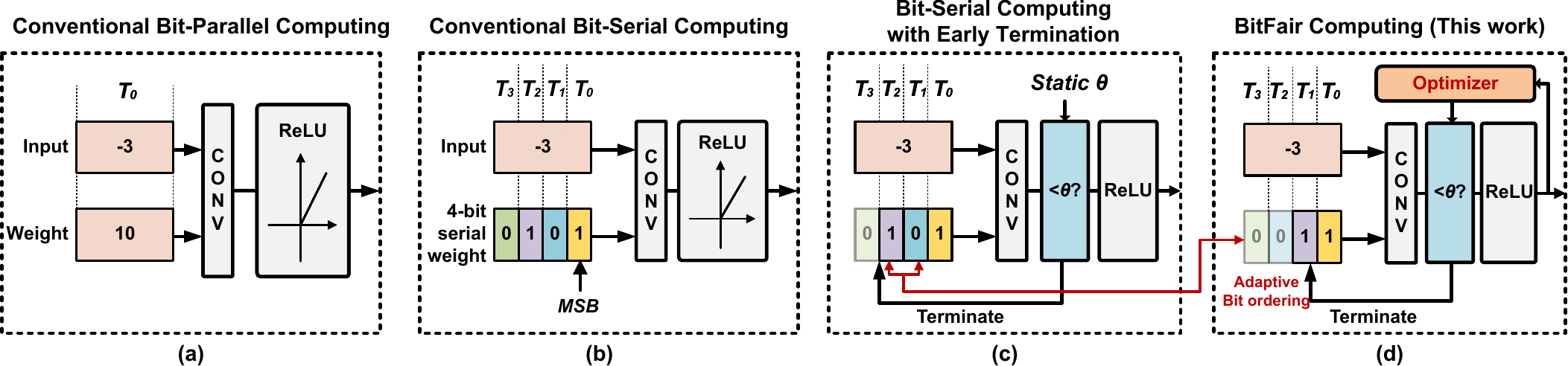}
    \caption{Execution paths for a ReLU-enabled convolution layer with different computing paradigms. (a) Conventional bit-parallel MAC units compute all bits in parallel. (b) Bit-serial processing consumes one bit per cycle. (c) Bit-serial processing with MSB-first early termination skips remaining bits once the partial sum is predicted to be negative. The fixed threshold $\theta$ is usually initialized from the normalization parameters or bias values and keeps static during inference. (d) Our proposed BitFair augments early termination with learnable thresholds and adaptive bit ordering to maximize skipped efficiency while preserving accuracy.}
    \label{fig:core_arch}
\end{figure*}

\subsection{Bit-Serial Neural Network Acceleration}

Bit-serial computing has emerged as a promising paradigm for energy-efficient neural network inference~\cite{judd2016stripes, albericio2017bit, sharify2019laconic, lee2018unpu}. Unlike conventional parallel processing that consumes all bits simultaneously (Figure~\ref{fig:core_arch}(a)), bit-serial architectures process quantized values sequentially, one bit at a time (Figure~\ref{fig:core_arch}(b)). This enables fine-grained control, exploiting bit-level sparsity in activations or weights. For example, PRAGMATIC~\cite{albericio2017bit} exploits bit-level sparsity by skipping computation cycles corresponding to zero bits in the binary representation. BitWave~\cite{shi2024bitwave} and BitPattern~\cite{wang2025bitpattern} further optimize bit-serial processing through structuring the bit-level sparsity to achieve higher energy efficiency.

However, these bit-serial accelerators only focus on static bit-level sparsity, i.e., zero bits already present in the quantized representation, without considering the runtime dynamic sparsity introduced by activation functions. In CNNs, ReLU activation functions clamp negative outputs to zero, inducing activation sparsity exceeding 50\% in deep layers~\cite{wen2016learning}. Conventional bit-serial accelerators fail to exploit \changed{the ReLU-induced sparsity}, processing all non-zero bits without considering the sign of the final accumulated value. This leaves substantial optimization opportunities unexplored, particularly for layers where many intermediate computations produce negative values that ReLU will ultimately zero out.

\subsection{Early Termination in CNNs}

While \acp{SNN} achieve conditional computation implicitly through the integrate-and-fire mechanism, CNN-based approaches can approximate a similar effect by predicting and skipping computations whose outputs will be zeroed by ReLU. The data-dependent sparsity introduced by ReLU activation functions during runtime, which is one form of dynamic sparsity~\cite{zhou2025exploiting}, brings significant opportunities for computational and memory access savings.

SnaPEA~\cite{akhlaghi2018snapea} computes partial sums in a value-based manner and skips the remaining weight values based on prediction results. When the accumulated partial sum indicates the final output will be negative, SnaPEA terminates the computation for that output position. While effective at reducing computation, SnaPEA's value-level granularity limits its ability to exploit finer-grained optimization opportunities within the binary representation of quantized values. To enable early termination within bit-serial processing, several works have explored bit-level prediction mechanisms. Song et al.~\cite{song2018prediction} propose a two-stage predict-and-execute approach that processes weights in MSB-first order. In the prediction stage, a manually determined number of MSBs is computed to predict whether the final result will be negative. If predicted negatively, the remaining LSBs are skipped. Similarly, PredictiveNet~\cite{lin2017predictivenet} employs MSB-first bit-serial processing with empirically determined values for the number of MSBs required to reliably predict final results and perform early termination. BitSET~\cite{pan2023bitset} further advances bit-serial early termination by introducing predefined thresholds derived from batch normalization parameters and dedicated encoding schemes to facilitate early termination decisions.

However, these bit-level approaches have fundamental limitations as shown in Figure~\ref{fig:core_arch}(c). First, MSB-first ordering may be suboptimal for layers where weight distributions do not align with positional bit significance. Second, these methods rely on manually determined or empirically fixed thresholds that cannot adapt to layer-specific characteristics, leading to suboptimal early termination decisions. \rev{Third, BitSET~\cite{pan2023bitset} makes early predictions reliable under a fixed threshold by moving away from the standard signed-integer format. It uses a custom weight encoding in which every bit contributes $\pm 2^{i}$. Although this encoding is a one-to-one mapping of the quantized values, it changes the numerical meaning of each weight bit. As a result, the datapath needs dedicated encoding logic, and the network has to undergo BitSET-aware fine-tuning until the weight distribution is compatible with the encoding. This couples early termination to both a non-standard representation and a specialized retraining procedure. These limitations motivate BitFair, shown in Figure~\ref{fig:core_arch}(d), which combines learnable thresholds with adaptive bit ordering to improve early-termination efficiency while preserving accuracy, without changing the underlying numerical format.}

\section{BitFair Algorithm}
\label{sec:bitfair_algorithm}
\subsection{Bit-Serial Processing with Early Termination}

\rev{For a quantized $N$-bit sign-magnitude weight $w$, with one sign bit $s$ and $N-1$ magnitude bits,} the value can be decomposed as:
\begin{equation}
w = s \times \sum_{j=0}^{\rev{N-2}} b_j \times 2^{j}
\end{equation}
where $b_j \in \{0, 1\}$ represents the bit at position $j$, and $s \in \{-1, +1\}$ is the sign bit.

\subsubsection{Conventional Bit-Serial Processing}
For a CNN layer, let $\mathcal{I} = \{i_0, i_1, \ldots, i_{M-1}\}$ denote the input positions contributing to an output, where $M = KH \times KW \times ICH$ for kernel dimensions $KH, KW$ and input channels $ICH$. The output $y$ of this layer \revcg{with bit-serial weight processing and bit-parallel activations} is:
\begin{equation}
y = \sum_{i=0}^{M-1}\sum_{j=0}^{\rev{N-2}} \left[ 2^{j} \times s_i \times b_{i,j} \times a_i \right] + \rev{\tilde{b}}
\label{eq:conventional_bit_serial_processing}
\end{equation}
\rev{where $a_i$ is the signed bit-parallel activation for input position $i \in \mathcal{I}$, $b_{i,j}$ is the weight bit at position $j$ for the $i$-th input position. $s_i$ is the sign bit of the sign-magnitude weight, and $\tilde{b}$ is the bias term.}

Equation~\ref{eq:conventional_bit_serial_processing} shows that conventional bit-serial implementations process the inner sum over bits $j$ first, then accumulate across inputs $i$. While computationally straightforward, this ordering prevents bit-level early termination, as partial sums aggregate only the processed inputs, providing limited information about the final output's sign. Moreover, while MSB-first processing accounts for positional bit significance, it overlooks the actual distribution of weight bits within a layer. When LSBs dominate substantially over MSBs, this ordering becomes inefficient, leading to \textit{``unfair''} bit treatment.

\subsubsection{BitFair's Bit-Level Early Termination}
To incorporate early termination into bit-serial processing, BitFair first reverses the computation order, processing each bit position across all inputs before moving to the next bit. A custom ordering $\omega$ is applied to define the bit processing sequence, instead of using the default MSB-first order. In this way, BitFair achieves \textit{``bit-fair''} processing by prioritizing the most informative bits, rather than relying solely on their inherent positional significance.

On this basis, an early termination decision can be made at bit-level granularity. Let $P_k$ represent the partial sum after processing $k$ bits under a given ordering $\omega$:
\begin{equation}
P_k = \sum_{j=0}^{k}\sum_{i=0}^{M-1}  \left[ 2^{\omega(j)} \times s_i \times b_{i,\omega(j)} \times a_i \right]
\label{eq:partial_sum}
\end{equation}
\rev{When $k = N-2$, all magnitude bits are processed.} For convolutional layers followed by ReLU activation, early termination can occur at $k < \rev{N-2}$ when the partial sum $P_k$ reliably predicts that the final output will be negative and thus zeroed by ReLU. Specifically, early termination occurs at step $k$ when:
\begin{equation}
P_k  \leq \theta^l
\end{equation}
where $\theta^l$ is BitFair's learned layer-specific threshold. 
\revcg{The threshold $\theta^l$ is defined on this bias-free partial sum and absorbs the layer bias or fused BN affine parameters, as detailed in Section~\ref{subsec:learnable_threshold}.}
\revcg{When this condition holds, BitFair predicts that the final output will be zeroed by ReLU and terminates the remaining bit-plane processing for that output. The threshold is learned to make this prediction accurate while trading off accuracy and saved bit-cycles.}

\subsection{Learnable Per-Layer Thresholds}
\label{subsec:learnable_threshold}

Conventional methods such as BitSET~\cite{pan2023bitset} employ fixed thresholds initialized from \ac{BN} layer parameters or bias values that remain static during inference. For a convolutional layer fused with a \ac{BN} layer, the threshold $\theta_0^l$ is initialized from the \ac{BN} parameters:
\begin{equation}
\label{eq:bn_threshold}
\theta_0^l = \left(-\frac{\beta}{\gamma}\right) \times \sqrt{\sigma^2 + \epsilon} + \mu
\end{equation}
where $\mu$, $\sigma$, $\gamma$, $\beta$, and $\epsilon$ are the mean, variance, scale, shift, and small constant parameters of the BN layer, respectively. For fully connected layers or convolutional layers without a subsequent BN layer, the threshold simplifies to $\theta_0^l = -\tilde{b}$, which can be derived as a special case of Equation~\ref{eq:bn_threshold} by setting $\mu = 0$, $\sigma = 1$, $\gamma = 1$, and $\beta = \tilde{b}$. However, fixed thresholds fail to adapt to varying input statistics, likely leading to suboptimal early termination.

\subsubsection{Threshold Parameterization}
BitFair uses the same initialization but makes thresholds trainable. For each layer $l$, we maintain a learnable threshold parameter $\theta^l$ initialized as:
\begin{equation}
\theta^l = \theta_0^l + \theta_x^l
\end{equation}
where $\theta_0^l$ is computed using Equation~\ref{eq:bn_threshold} for layers with BN or $\theta_0^l = -\tilde{b}_l$ for layers without BN, and $\theta_x^l$ is a learnable offset parameter.

\subsubsection{Differentiable Soft Thresholding}
The direct application of hard threshold comparisons creates non-differentiable decision points that prevent gradient flow. To enable end-to-end training, we use a temperature-controlled sigmoid gate. At bit position $k$, the early-termination gate is:
\begin{equation}
\rev{\mathbf{G}^l_k} = \sigma\left(-\frac{\rev{\mathbf{P}^l_k} - \theta^l}{T}\right)
\label{eq:soft_gate}
\end{equation}
where $\sigma(\cdot)$ denotes the sigmoid function, \rev{$\mathbf{P}^l_k$ represents the stacked partial sums $P_k$ across all outputs at layer $l$}, and $T$ is the temperature parameter. This formulation directly relaxes the hard early termination condition $\rev{\mathbf{P}^l_k} \leq \theta^l$: when $\rev{\mathbf{P}^l_k} - \theta^l \leq 0$, we obtain high $\rev{\mathbf{G}^l_k} \approx 1$ (early termination likely); conversely, when $\rev{\mathbf{P}^l_k} - \theta^l > 0$, we obtain low $\rev{\mathbf{G}^l_k} \approx 0$ (continue to next bit).

\begin{figure}[t]
    \centering
    \includegraphics[width=0.90\linewidth]{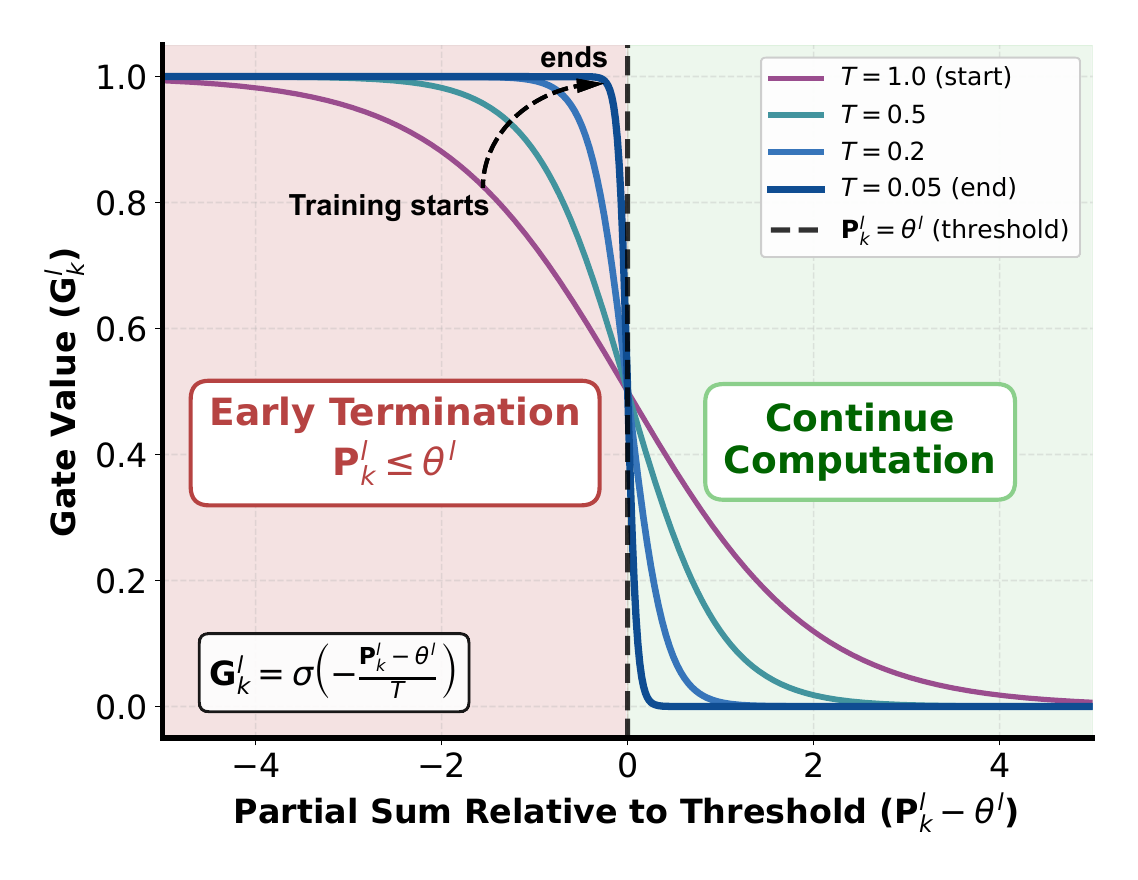}
    \caption{\rev{Effect of temperature annealing on sigmoid gate sharpness: high temperatures enable soft decisions during training, while low temperatures produce hard decisions approaching the step function used at inference.}}
    \label{fig:temperature_annealing}
\end{figure}

\revcg{A single gate relaxes one termination decision. Because the decisions are sequential and the output is zeroed as soon as the \emph{first} gate fires, we compose the per-position gates into a prefix-product survival probability. Let $K=N_l-2$ denote the final magnitude-bit position of layer $l$ and define}
\begin{equation}
\label{eq:multi_bit_soft_output}
\revcg{
\mathbf{S}^l_k=\prod_{m=0}^{k}(1-\mathbf{G}^l_m),
\quad k=0,\ldots,K.
}
\end{equation}
\revcg{Here, $\mathbf{S}^l_k$ is the soft probability that no early-termination gate has fired through bit position $k$. Any fired gate drives this product toward zero and suppresses the output. Thus, only outputs that survive through the final bit position contribute to the differentiable training output:}
\begin{equation}
\label{eq:soft_output}
\revcg{\hat{\mathbf{A}}^l = \mathbf{S}^l_{K} \odot \mathrm{ReLU}(\mathbf{P}^l_K + \tilde{\mathbf{b}}^l)}
\end{equation}
\revcg{During inference, the soft gates are replaced by the hard sequential comparison $\mathbf{P}^l_k \leq \theta^l$. The first satisfying bit position terminates the output and writes zero. If no comparison fires through $k=K$, the layer emits the conventional full-bit output.}
\subsubsection{Training Objective}
\rev{Equation~\ref{eq:soft_output} casts each early-termination layer $l\in\mathbb{L}$ as a differentiable transfer function $\hat{\mathbf{A}}^l = f^l(\hat{\mathbf{A}}^{l-1};\mathbf{W}^l,\theta^l)$, where $\mathbf{W}^l$ is the layer's quantized $N$-bit sign-magnitude weight matrix. In a neural network model, composing all $L$ layers propagates the input $\mathbf{x}$ to the output logits,}
\begin{equation}
\label{eq:network_forward}
\rev{\hat{\mathbf{z}} = f^{L}\!\circ f^{L-1}\!\circ \cdots \circ f^{1}(\mathbf{x}), \qquad \hat{\mathbf{A}}^0 = \mathbf{x}}
\end{equation}
\rev{so each threshold $\theta^l$ influences the prediction through the smooth survival probability $\mathbf{S}^l$ of Equation~\ref{eq:multi_bit_soft_output}. Hence the logits $\hat{\mathbf{z}}$ are differentiable with respect to all thresholds $\{\theta^l\}_{l\in\mathbb{L}}$, which lets us train them end-to-end.}
\rev{The thresholds are trained with the standard task loss plus a bit-position survival regularizer. For clarity, the loss below is written for a single batch input. For logits $\hat{\mathbf{z}}$ and labels $\mathbf{t}$, the total loss takes the conventional form}
\begin{equation}
\label{eq:threshold_training_loss}
\rev{
\mathcal{L}
= \mathcal{L}_{\mathrm{CE}}(\hat{\mathbf{z}}, \mathbf{t})
+ \lambda_{\mathrm{bit}}\, \mathcal{L}_{\mathrm{bit}}
}
\end{equation}
\rev{where $\mathcal{L}_{\mathrm{CE}}$ is the cross-entropy task loss and $\lambda_{\mathrm{bit}}$ balances accuracy against efficiency. The regularizer $\mathcal{L}_{\mathrm{bit}}$ penalizes high survival probabilities across magnitude-bit positions in the early-termination layers $\mathbb{L}$,}
\begin{equation}
\label{eq:bit_regularizer}
\revcg{
\mathcal{L}_{\mathrm{bit}}
= \frac{1}{|\mathbb{L}|}
\sum_{l\in\mathbb{L}}
\frac{1}{|\mathcal{U}^l|}
\sum_{u\in\mathcal{U}^l}
\frac{1}{N_l-1}
\sum_{k=0}^{N_l-2}
\mathbf{S}_{k}^{l,u}
}
\end{equation}
\revcg{where $\mathcal{U}^l$ is the index set of output elements in layer $l$, $\mathbf{S}_k^{l,u}$ is the survival probability of output element $u$ after processing bit position $k$, and $N_l-1$ is the number of magnitude-bit positions. The inner sum $\sum_{k}\mathbf{S}_{k}^{l,u}$ gives the expected number of magnitude-bit positions processed for output $u$. Normalizing by $N_l-1$ gives the expected processed-bit fraction. Minimizing $\mathcal{L}_{\mathrm{bit}}$ therefore encourages earlier termination, while $\mathcal{L}_{\mathrm{CE}}$ preserves task accuracy.}

\subsubsection{Temperature Annealing}
To transition from soft decisions during training to hard decisions at inference, we anneal the temperature $T$ using an exponential decay schedule:
\begin{equation}
\label{eq:temp_anneal}
T(e) = T_{0} \times \left(\frac{T_{E}}{T_0}\right)^{e / E}
\end{equation}
where $e$ is the current epoch and $E$ is the total training epochs. We decay $T$ from $T_0 = 1.0$ to $T_E = 0.05$ over training. At high temperatures (early training), the sigmoid gate remains smooth, yielding soft early-termination decisions that support stable end-to-end optimization of $\theta^l$. As $T$ decreases, the gate progressively sharpens, approaching the hard comparison $\rev{\mathbf{P}^l_k} \leq \theta^l$ used during inference. Consequently, the model learns thresholds under a differentiable relaxation while converging toward the discrete early-termination behavior deployed at inference time. Intuitively, a larger $T$ prevents the gate from saturating too early, enabling thresholds to adapt to evolving activation statistics. As annealing progresses, lower $T$ reduces ambiguity around the boundary and makes early termination more decisive, minimizing train--test mismatch when switching to hard comparisons at inference.

\begin{algorithm}[t]
    \caption{Greedy Bit Ordering Search}
    \label{alg:bit_ordering}
    \begin{algorithmic}[1]
    \REQUIRE Model $\mathcal{M}$, calibration dataset $\mathcal{D}_{\text{cal}}$, thresholds $\{\theta^l\}$, baseline accuracy $\text{Acc}_{\text{baseline}}$
    \ENSURE \revcg{Greedy-selected bit ordering $\omega^l$ for each layer $l$}
    \FOR{each layer $l$ in $\mathcal{M}$}
        \STATE $\text{selected\_order} \leftarrow []$ \COMMENT{Initialize empty ordering}
        \STATE $\text{remaining\_bits} \leftarrow [N-2, N-3, \ldots, 1, 0]$
        \FOR{each ordering slot $j$ from $0$ to $N-2$}
            \STATE $\text{best\_score} \leftarrow 0$
            \STATE $\text{best\_bit} \leftarrow \text{NULL}$
            \FOR{each candidate bit $b$ in $\text{remaining\_bits}$}
                \STATE $\text{test\_order} \leftarrow \text{selected\_order} + [b]$ \COMMENT{Fill the remaining bits by MSB-first order}
                \STATE $\text{ETR} \leftarrow \text{EvaluateETR}(\mathcal{M}, \mathcal{D}_{\text{cal}}, \text{test\_order}, \theta^l)$
                \revcg{
\STATE $\text{Acc} \leftarrow \text{EvaluateAcc}(\mathcal{M}, \mathcal{D}_{\text{cal}}, \text{test\_order}, \theta^l)$
\STATE $\text{AccLoss} \leftarrow \text{Acc}_{\text{baseline}}-\text{Acc}$
\STATE $\text{score} \leftarrow \text{ETR} / (\text{AccLoss}+\epsilon)$}
                \IF{$\text{score} > \text{best\_score}$}
                    \STATE $\text{best\_score} \leftarrow \text{score}$
                    \STATE $\text{best\_bit} \leftarrow b$
                \ENDIF
            \ENDFOR
            \STATE $\text{selected\_order}.\text{append}(\text{best\_bit})$
            \STATE $\text{remaining\_bits}.\text{remove}(\text{best\_bit})$
        \ENDFOR
        \STATE $\omega^l \leftarrow \text{selected\_order}$
    \ENDFOR
    \RETURN $\{\omega^l\}$ for all layers
    \end{algorithmic}
\end{algorithm}

\subsection{Adaptive Bit Ordering Optimization}
To maximize the effectiveness of early termination, an appropriate bit ordering strategy is crucial. While MSB-first ordering processes large-magnitude bits first, layer-specific weight distributions may benefit from alternative orderings prioritizing bits with the highest discriminative power. BitFair employs \changed{a} greedy search strategy to discover \changed{an effective} bit ordering for each layer.

\subsubsection{Problem Formulation}
For $N$-bit weights in sign-magnitude representation, the processing order of the $N-1$ magnitude bits needs optimization. 
\revcg{We seek an ordering $\omega^l=[\omega^l(0),\omega^l(1),\ldots,\omega^l(N-2)]$, where $\omega^l(j)$ denotes the original magnitude-bit position processed at ordering slot $j$.}
The optimization objective balances \ac{ETR} against accuracy preservation:
\begin{equation}
\label{eq:score}
\revcg{
\text{score}(\omega)
=
\frac{\text{ETR}(\omega)}
{\text{AccLoss}(\omega)+\epsilon}
}
\end{equation}
\revcg{where $\text{ETR}(\omega)$ measures the early-termination rate under ordering $\omega$, $\text{AccLoss}(\omega)=\text{Acc}_{\text{baseline}}-\text{Acc}(\omega)$ is the calibration accuracy loss, and $\epsilon$ is a small constant to prevent division by zero. Higher scores indicate orderings achieving high termination rates with minimal accuracy loss.}

\subsubsection{Greedy Search Procedure}
Algorithm~\ref{alg:bit_ordering} presents the proposed greedy search approach. Starting with an \changed{empty} ordering, we iteratively select one bit at a time over the N-1 positions. At each position $k$, all remaining unselected bit positions are candidates for the next position in the ordering. A \textit{test\_order} is constructed by appending the candidate bit $b$, then completing the remaining unselected bit positions in MSB-first order, ensuring a complete (N-1)-bit ordering before evaluation.

\revcg{For each \textit{test\_order}, we evaluate the early-termination rate and calibration accuracy by running inference on $\mathcal{D}_{\text{cal}}$ with the current layer-specific threshold $\theta^l$. The candidate with the largest score in Equation~\ref{eq:score} is selected. This process repeats until all $N-1$ magnitude-bit positions are ordered.}

\rev{Greedy search is proven effective in practice when dealing with neural-network layer-wise optimization problems~\cite{li2022bitcluster, akhlaghi2018snapea}. For our problem, given $N-1$ magnitude bits, instead of exhaustively searching all $(N-1)!$ permutations, the greedy approach uses $N-1$ iterations. At ordering slot $j$, it evaluates $N-1-j$ candidates, resulting in a total of $N(N-1)/2$ evaluations. This makes the search tractable while still capturing the dominant benefits of bit ordering. While more sophisticated search methods such as reinforcement learning could potentially find further gains, the current greedy search already provides significant improvements over the fixed MSB-first baseline with limited offline cost, as shown in Figure~\ref{fig:impact_abo}. A broader exploration of the ordering search space is left to future work.}

\begin{figure*}[t]
    \centering
    \includegraphics[width=0.90\linewidth]{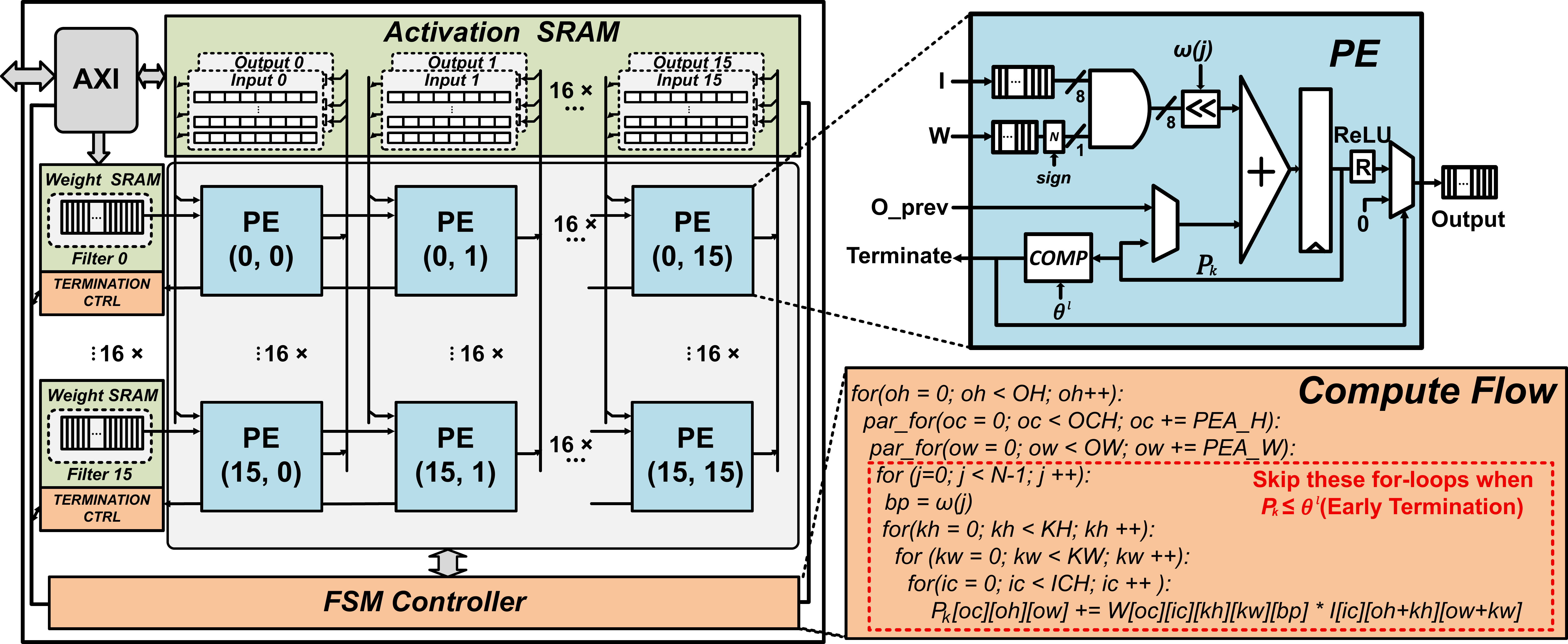}
    \caption{\rev{The BitFair accelerator architecture: (left) a 16$\times$16 PE array with its activation and weight memory hierarchy, AXI interface, and FSM controller; (top right) the PE microarchitecture, featuring a bit-serial datapath and early-termination logic; (bottom right) the nested-loop compute flow of the architecture.}}
    \label{fig:accelerator_arch}
\end{figure*}

\section{BitFair Hardware Architecture}
\label{sec:hardware}

\subsection{Architecture Overview}
Figure~\ref{fig:accelerator_arch} shows the BitFair accelerator. An AXI interface configures layer dimensions, bit ordering, thresholds, and loads weights and activations into on-chip SRAM. A 16$\times$16 \ac{PE} array exploits parallelism over output channels and spatial positions, with column-shared input buffers, per-filter weight banks, and local output buffers aggregated into a global output memory. The distributed termination controllers enable independent early termination per PE and allow clock-gating when new data is not yet available. A lightweight \ac{FSM} controller coordinates the system operation by managing the bit-serial processing flow and aggregating early-termination signals from all \acp{PE} to control data movement between memory and the \ac{PE} array.

\subsection{Processing Element Microarchitecture}
Each \ac{PE} integrates bit-serial computation with early termination capability. The datapath receives a weight bit ($W$), an 8-bit input activation ($I$), and a bit-position significance $\omega(j)$, where $\omega(j)$ maps the current bit position $j$ to its corresponding significance. The AND gate computes the partial product between the weight bit and \changed{the} activation, with sign handling logic that derives the partial product sign from the sign bits of both operands. A barrel shifter scales the result by $2^{\omega(j)}$, implementing the bit-position significance according to the layer-specific ordering rather than the fixed MSB-first sequence. At the end of the datapath, an accumulator maintains the running partial sum $P_k$ across bit positions. It receives the current bit-level product from the multiplier and the previous partial sum from the feedback path.

The early termination logic compares the partial sum $P_k$ against the layer-specific threshold using a comparator. When $P_k \leq \theta^l$, the terminate signal is asserted, halting further bit processing for the current output. A multiplexer selects between the accumulated partial sum (normal completion) or zero (early termination) before applying ReLU activation to produce the final output.

\subsection{PE Array Organization}

The 16$\times$16 PE array follows an output-stationary dataflow, where each PE accumulates partial sums for a specific output activation. Input activations are broadcast to rows of PEs, while weights are distributed to columns, minimizing partial sum movement throughout the accumulation process.

The output-stationary approach is particularly advantageous for BitFair's early termination mechanism. Since each PE independently accumulates a single output value, early termination decisions are made locally without coordination across PEs. When a PE detects that its output will be negative, it halts immediately without affecting neighboring PEs processing different outputs. This distributed decision-making eliminates global synchronization overhead and enables fine-grained per-output efficiency gains.

\subsection{FSM Controller}
At the start of each layer, the \ac{FSM} controller loads layer-specific parameters from configuration registers, including the bit ordering $\omega^l$, threshold $\theta^l$, and layer dimensions (channels, spatial size, and kernel size). It then initializes the address generators and schedules the movement of activations and weights between on-chip SRAM buffers and the 16$\times$16 \ac{PE} array. 

During execution, the controller orchestrates the nested loops of the convolution and advances the bit-serial computation by broadcasting the current bit-select and shift-amount signals corresponding to $\omega^l$ to all \acp{PE}. In each bit step, it collects early-termination indications from the array and uses them to suppress further work for outputs that have already converged to zero after ReLU, while allowing other \acp{PE} to continue. The controller manages write-back of completed outputs to the global output memory and triggers the next output tile or channel group once resources are available. To maintain correctness under bandwidth limits, it can stall the compute pipeline when operand data is not ready, and it enables clock-gating opportunities for idle \acp{PE} and buffers to reduce wasted switching activity.

\subsection{Compute Flow}
Figure~\ref{fig:accelerator_arch} also shows the nested loop structure of the compute flow adopted in the BitFair accelerator controller. The flow follows an output-stationary dataflow pattern, where each \ac{PE} accumulates partial sums for an output activation. To compute the partial sum in a bit-serial manner, we introduce an extra loop over weight bit positions, since each PE processes weights bit-serially. The termination control logic continuously monitors these partial sums, comparing them against the layer threshold. When the early termination condition is satisfied within a \ac{PE}, that \ac{PE} halts computation for the current output and notifies the controller, effectively skipping the remaining unprocessed bits and saving energy.

\section{\rev{Results}}
\label{sec:evaluation}

\subsection{\rev{Experimental} Setup}

\subsubsection{Algorithm}
To validate the computational performance and efficiency of BitFair on edge vision tasks, we employ VGGNet~\cite{simonyan2014very}-like network architectures with a CONV-BN-ReLU structure. All networks are trained and evaluated using the PyTorch~\cite{paszke2019pytorch} framework. \rev{Table~\ref{tab:format} summarizes the numerical formats used in our experiments: 8-bit 2's-complement activations, 8-bit sign-magnitude weights (one sign bit and seven magnitude bits), and 16-bit partial sums in 2's-complement format. The 8-bit weight setting corresponds to the maximum configured precision in our evaluation while the same weight-bit-serial datapath also supports 1$\sim$7 serial magnitude bits plus a sign bit.}

\begin{table}[t]
    \centering
    \rev{%
    \caption{\rev{Numerical format summary for BitFair.}}
    \label{tab:format}
    \small
    \setlength{\tabcolsep}{4pt}
    \begin{tabular*}{\columnwidth}{@{\extracolsep{\fill}}lcc@{}}
    \toprule
    \textbf{Operand} & \textbf{Representation} & \textbf{Precision (bits)} \\
    \midrule
    Activation                      & 2's complement & 8 \\
    \addlinespace
    Weight\textsuperscript{1}       & sign-magnitude & 8 \\
    \quad sign bit                  & -              & 1 \\
    \quad magnitude bits            & unsigned       & 7 \\
    \addlinespace
    Partial sum                     & 2's complement & 16 \\
    \bottomrule
    \end{tabular*}\\[2pt]
    \footnotesize
    \noindent\parbox{\columnwidth}{\raggedright
    \rev{\textsuperscript{1} Experimental results use 8-bit weights. BitFair's weight-bit-serial datapath also supports 1$\sim$7 serial magnitude bits plus a sign bit.}}
    }
\end{table}

Table~\ref{tab:network_specs} summarizes the evaluated network architectures and specifications for different datasets. \rev{It also lists the input type, input size, normalization setting, and number of event bins for event datasets. Frame datasets are processed as single static images. For the event datasets, N-MNIST and DVSGesture, raw events are integrated into 16 event bins using the equal-event-count partitioning method from SpikingJelly~\cite{fang2023spikingjelly}: each bin accumulates an equal share of the recording's total events, rather than spanning a fixed time interval, so it holds on average about 262 events per bin for N-MNIST and 19.8K for DVSGesture.}

\revcg{To implement BitFair's bit-level early-termination algorithm, we decompose weights into sign-magnitude representation and proceed in three stages. First, we train the quantized baseline and initialize layer-wise thresholds $\theta^l$ using Equation~\ref{eq:bn_threshold} from the fused BN or bias parameters, with learnable offsets $\theta_x^l$ initialized to zero. Second, using the default MSB-first order, we optimize the thresholds end-to-end as described in Section~\ref{subsec:learnable_threshold}, with temperature annealing from $T=1.0$ to $T=0.05$. Third, we fix the learned thresholds and run the \ac{ABO} search in Algorithm~\ref{alg:bit_ordering} on the calibration set to determine the final layer-wise bit-processing order. This final ABO step is performed once per trained model and dataset and does not affect inference-time latency or energy.}

\begin{table*}[t]
    \centering
    \caption{BitFair's network specifications for different datasets.}
    \small
    \setlength{\tabcolsep}{4pt}
    \resizebox{\textwidth}{!}{%
    \begin{tabular}{llccccccccc}
    \toprule
    \textbf{Dataset} & \rev{\textbf{Task}} & \rev{\makecell[c]{\textbf{Input}\\\textbf{type}}} & \rev{\makecell[c]{\textbf{Input}\\\textbf{size}\textsuperscript{1}}} & \rev{\makecell[c]{\textbf{Norm.}\textsuperscript{3}}} & \rev{\makecell[c]{\textbf{Event}\\\textbf{bins}}} & \rev{\makecell[c]{\textbf{Calib.}\\\textbf{set}\textsuperscript{2}}} & \makecell[c]{\textbf{Conv-BN-ReLU}\\\textbf{Config. (Channels)}} & \makecell[c]{\textbf{\#Params}} & \makecell[c]{\textbf{Acc.}\\\textbf{(Float)}} & \makecell[c]{\textbf{Acc.}\\\textbf{(BitFair)}} \\
    \midrule
    N-MNIST~\cite{orchard2015converting-nmnist}     & \rev{Handwritten digit} & \rev{event} & \rev{34$\times$34$\times$2}   & \rev{None} & \rev{16} & \rev{8K} & 16–16–32–32      & 9.9K  & 98.43\% & 97.72\% \\
    DVSGesture~\cite{amir2017low-dvs-gesture}       & \rev{Hand gesture}      & \rev{event} & \rev{128$\times$128$\times$2} & \rev{None} & \rev{16} & \rev{256} & 8–16–16–32–32    & 18.5K & 97.40\% & 96.53\% \\
    MNIST~\cite{lecun2002gradient-mnist}            & \rev{Handwritten digit} & \rev{frame} & \rev{28$\times$28$\times$1}   & \rev{Mean/std} & \rev{-}  & \rev{8K} & 8–16–32–32       & 8.6K  & 98.66\% & 97.46\% \\
    SVHN~\cite{netzer2011reading-svhn}              & \rev{Street-view digit} & \rev{frame} & \rev{32$\times$32$\times$3}   & \rev{Mean/std} & \rev{-}  & \rev{8K} & 16–16–16–32–64   & 28.9K & 93.35\% & 91.83\% \\
    \rev{\acs{VWW}~\cite{chowdhery2019visual}}   & \rev{Person detection}  & \rev{frame} & \rev{96$\times$96$\times$3}   & \rev{None} & \rev{-} & \rev{12K} & \rev{8–16–28–30–32–34} & \rev{31.7K} & \rev{87.84\%} & \rev{86.52\%} \\
    \bottomrule
    \end{tabular}%
    }
    \par
    \vspace{2pt}
    \begin{minipage}{\textwidth}
    \footnotesize
    \rev{\textsuperscript{1}Input image size (height $\times$ width $\times$ channels). For \acs{VWW}, images are downscaled from their original resolution.}\\
    \rev{\textsuperscript{2}Calibration-set size used for the adaptive bit-ordering search.}\\
    \rev{\textsuperscript{3}Mean/std denotes standard per-channel normalization with training-set statistics.}
    \end{minipage}
    \label{tab:network_specs}
\end{table*}

\subsubsection{Hardware}
We implement the BitFair accelerator in Verilog HDL and synthesize it using Cadence Genus, followed by place-and-route in Cadence Innovus using the GlobalFoundries 12~nm FinFET technology. We evaluate the design across a range of supply voltages and clock frequencies, from 0.55 to 0.80~V and from 20 to 500~MHz, respectively. Power analysis is performed using Synopsys PrimePower with post-layout parasitic extraction from Synopsys StarRC and realistic activity factors obtained from functional simulations in Synopsys VCS.

\subsection{Accuracy and Efficiency}
\begin{figure}[t]
    \centering
    \includegraphics[width=\linewidth]{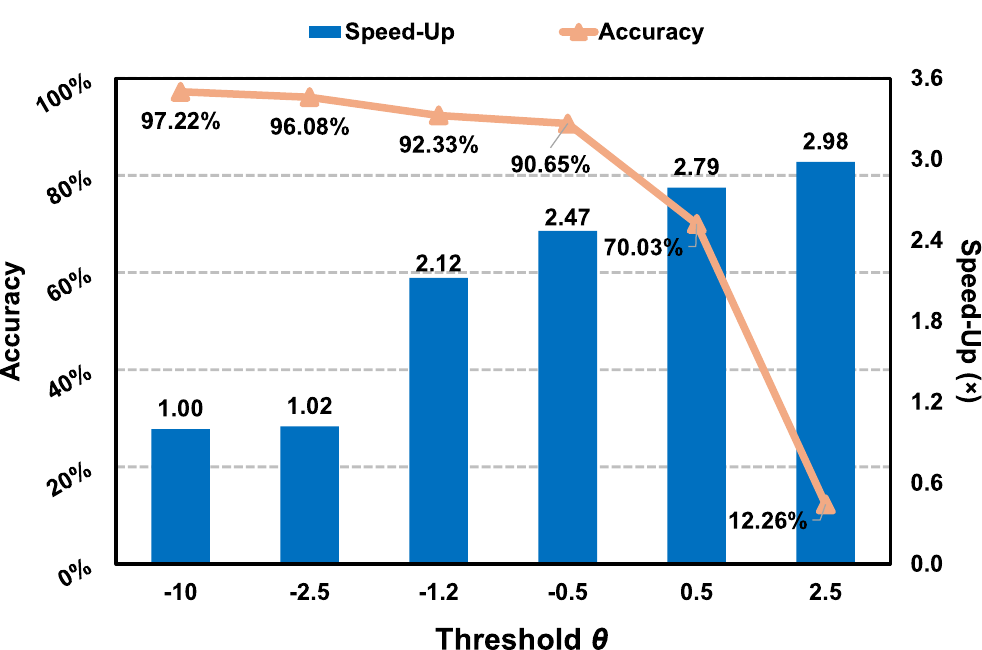}
    \caption{\rev{Ablation study of the learnable early-termination threshold: accuracy and efficiency vs.\ model-wise threshold values on DVSGesture~\cite{amir2017low-dvs-gesture} dataset. 
    Here, ``speed-up'' is the ratio between the vanilla 8-bit bit-serial compute latency and the BitFair compute latency under the same layer schedule, computed from bit-serial calculation cycles. 
    }
    }
    \label{fig:impact_ithres}
\end{figure}
\begin{figure}[t]
    \centering
    \includegraphics[width=\linewidth]{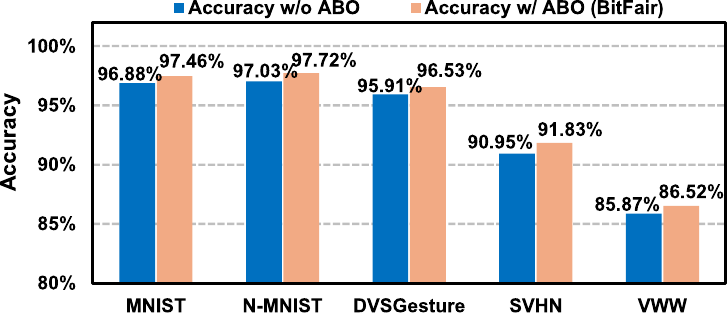}\\
    (a)\\[0.5em]
    \includegraphics[width=\linewidth]{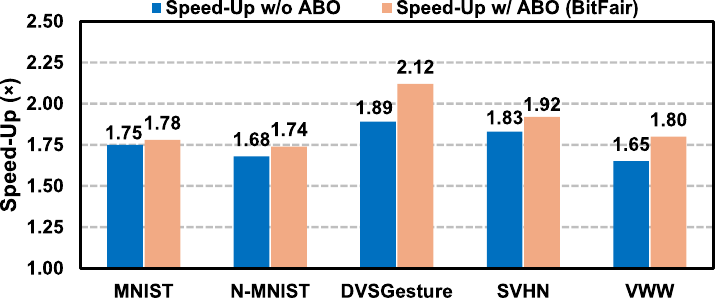}\\
    (b)
    \caption{\rev{Ablation study of adaptive bit ordering (ABO), comparing results with and without ABO across datasets: (a) accuracy and (b) speed-up.}}
    \label{fig:impact_abo}
\end{figure}

\begin{table*}[t]
    \centering
    \caption{Comparison of BitFair with state-of-the-art edge vision accelerators.}
    \label{tab:sota_comparison}
    \scriptsize
    \setlength{\tabcolsep}{2pt}
    \renewcommand{\arraystretch}{1.3}
    \resizebox{\textwidth}{!}{
    \begin{tabular}{lccccccccccccccccc}
    \toprule
     & & & & & & & & & & & \multicolumn{3}{c}{\textbf{N-MNIST}} & \multicolumn{3}{c}{\textbf{DVSGesture}} \\
    \cmidrule(lr){12-14} \cmidrule(lr){15-17}
    \textbf{Work} & \textbf{Tech} & \textbf{Voltage} & \textbf{Model} & \textbf{SRAM} & \textbf{Freq} & \textbf{Area} & \textbf{W-Prec.} & \textbf{Power} & \textbf{Eff.\textsuperscript{1}$\downarrow$} & \textbf{Eff.\textsuperscript{2}$\uparrow$} & \textbf{Acc$\uparrow$} & \textbf{FPS$\uparrow$} & \textbf{EDP$\downarrow$} & \textbf{Acc$\uparrow$} & \textbf{FPS$\uparrow$} & \textbf{EDP$\downarrow$} \\
     & (nm) & (V) &  & (KB) & (MHz) & (mm²) & (bits) & (mW) & (pJ/SOP) & (BTOPS/W) & (\%) & (FPS) & (nJ$\cdot$s) & (\%) & (FPS) & (nJ$\cdot$s) \\
    \midrule
    ISSCC'22~\cite{frenkel2022reckon} & 28 & 0.50 & SRNN & 138 & 13 & 0.45 & 8 & \textbf{$<$0.08} & 5.3 & \revfinal{3.0} & — & — & — & 87.3 & 1.67 & 27E3 \\
    ISSCC'23~\cite{zhang202322} & 28 & 0.56 & SFNN & 266.5 & 40 & 1.25 & 8 & 2.91 & 1.5 & \revfinal{10.6} & 96.0 & 8489 & 0.04 & 92.0 & 746 & 5.23 \\
    ISSCC'24~\cite{liu202430} & 22 & 0.55 & SCNN & — & 51 & 2.23 & 4/8 & 0.52 & 3.78 & \revfinal{4.2} & 97.0 & 137 & 27.7 & 94.0 & 20.2 & 1279 \\
    CICC'25~\cite{yang202540nm} & 40 & 0.72 & SNN & 161 & 60 & 1.42 & 4/8 & 4.63 & 0.62 & \revfinal{25.8} & 95.6 & \textbf{21K} & \textbf{0.009} & 90.5 & 2629 & 0.54 \\
    VLSI'25~\cite{fu2025neuc} & 130 & 0.9 & SNN & 128 & 2.5 & 6.75 & 4 & $<$0.1 & 0.28 & \revfinal{57.2} & 97.1 & — & — & 90.1 & — & — \\
    \rev{TCASI'24~\cite{fu2024593nj}} & \rev{65} & \rev{0.7} & \rev{CNN} & \rev{12} & \rev{0.5} & \rev{0.68} & \rev{8} & \rev{0.03} & \rev{—\textsuperscript{3}} & \rev{\revfinal{38.2}\textsuperscript{3}} & \rev{—} & \rev{—} & \rev{—} & \rev{92.4} & \rev{58.5} & \rev{10.15} \\
    \arrayrulecolor{gray}\midrule\arrayrulecolor{black}
    \multirow{4}{*}{This Work} & \multirow{4}{*}{12} & 0.55 & \multirow{4}{*}{CNN} & \multirow{4}{*}{104} & 40 & \multirow{4}{*}{\textbf{0.34}} & \multirow{4}{*}{\rev{8}} & 0.8 & 0.08 & \revfinal{204.8} & \multirow{4}{*}{\textbf{97.7}} & 1700 & 0.28 & \multirow{4}{*}{\textbf{96.5}} & 645 & 1.92 \\
     & & 0.55 & & & 200 & & & 3.5 & \textbf{0.07} & \revfinal{\textbf{234.0}} & & 8500 & 0.049 & & 3225 & 0.34 \\
     & & 0.65 & & & 400 & & & 9.4 & 0.10 & \revfinal{174.4} & & 17K & 0.033 & & 6448 & 0.23 \\
     & & 0.70 & & & \textbf{500} & & & 13.7 & 0.11 & \revfinal{149.6} & & \textbf{21K} & 0.030 & & \textbf{8060} & \textbf{0.21} \\
    \bottomrule
    \end{tabular}
    }
    \normalsize
    \vspace{-2mm}
    \raggedright
    \scriptsize
    \begin{minipage}{\textwidth}
    \textsuperscript{1}Power efficiency in pJ/SOP ($\downarrow$lower is better). SOP (Synaptic Operations Per Second) measures operations between a 1-bit spike and a $\text{\#Wbits}$ weight. This definition is equivalent to the weight-bit-serial operations on 8-bit activations used in this work. \\
    \textsuperscript{2}Power efficiency in BTOPS/W ($\uparrow$higher is better)\rev{~\cite{shi2025sparsecol}, where BTOPS (Binary Tera-Operations Per Second) is a bit-normalized throughput defined as $\text{BTOPS} = \text{\#Wbits} \times \text{\#Abits} \times \text{TOPS}$, where TOPS denotes Tera Operations per Second and \#Wbits, \#Abits are the weight and activation bit-precisions. For SNN-based works the activations are single-bit spikes ($\text{\#Abits}=1$), so this reduces to $\text{BTOPS} = \text{\#Wbits} \times \text{TSOPS}$, with TSOPS the Tera Spiking Operations per Second. For this work, the bit-serial datapath processes one weight bit per cycle ($\text{\#Wbits}=1$), so the metric reduces to $\text{BTOPS} = \text{\#Abits} \times \text{TOPS}$. \revfinal{Following~\cite{shi2025sparsecol}, one multiplication or one addition is counted as one operation; therefore, one multiply-accumulate operation corresponds to two operations.} For the bit-parallel TCASI'24, the full weight and activation precisions are used.
    \revcg{This bit-normalized metric provides a precision-aware reference across designs with different operand widths. Nevertheless, because CNN and SNN accelerators use different coding schemes and dataflows, we interpret it together with same-workload accuracy, FPS, and EDP rather than as a standalone absolute comparison.}
    } \\
    \textsuperscript{3}\rev{TCASI'24~\cite{fu2024593nj} is a bit-parallel CNN accelerator, so the pJ/SOP metric (defined for single-bit spike operations) does not apply. Its BTOPS/W is computed from the bit-normalized definition above using its reported model architecture and per-inference energy.}
    \end{minipage}
    \normalsize
\end{table*}
Table~\ref{tab:network_specs} summarizes BitFair's accuracy across edge vision tasks. Networks are first trained in floating-point, then quantized to 8-bit, and fine-tuned with BitFair's learnable thresholds and adaptive bit ordering. BitFair maintains high accuracy on all datasets, with at most 1.6\% degradation compared to the floating-point baselines.

\begin{table}[t]
    \centering
    \caption{Comparison of CNN early-termination methods across edge vision datasets.}
    \label{tab:cnn_et_comparison}
    \small
    \setlength{\tabcolsep}{3pt}
    \resizebox{\columnwidth}{!}{%
    \begin{tabular}{@{}l cc cc cc@{}}
    \toprule
    & \multicolumn{2}{c}{Vanilla$^{1}$} & \multicolumn{2}{c}{BitSET$^{2}$~\cite{pan2023bitset}} & \multicolumn{2}{c}{\textbf{BitFair (Ours)}} \\
    \cmidrule(lr){2-3} \cmidrule(lr){4-5} \cmidrule(lr){6-7}
    Dataset & Accuracy(\%) & Speed & Acc. & Spd. & Acc. & Spd. \\
    \midrule
    MNIST       & 98.23 & 1.00$\times$ & 96.44 & 1.59$\times$ & \textbf{97.46} & \textbf{1.78$\times$} \\
    N-MNIST     & 98.17 & 1.00$\times$ & 97.03 & 1.68$\times$ & \textbf{97.72} & \textbf{1.74$\times$} \\
    DVSGesture  & 97.11 & 1.00$\times$ & 95.20 & 1.96$\times$ & \textbf{96.53} & \textbf{2.12$\times$} \\
    SVHN        & 92.94 & 1.00$\times$ & 89.77 & 1.72$\times$ & \textbf{91.83} & \textbf{1.92$\times$} \\
    \rev{\acs{VWW}}   & \rev{87.17} & \rev{1.00$\times$} & \rev{85.81} & \rev{1.61$\times$} & \rev{\textbf{86.52}} & \rev{\textbf{1.80$\times$}} \\
    \bottomrule
    \end{tabular}%
    }
    \par
    \vspace{2pt}
    \begin{minipage}{\columnwidth}
    \footnotesize
    $^{1}$Vanilla denotes an 8-bit vanilla bit-serial baseline without early termination.\\
    $^{2}$BitSET results are reproduced following the methodology in~\cite{pan2023bitset}, as the original work did not evaluate on these datasets.
    \end{minipage}
\end{table}

\subsubsection{Impact of Threshold Values}
Figure~\ref{fig:impact_ithres} illustrates the accuracy--efficiency trade-off by sweeping a single model-wise threshold on DVSGesture~\cite{amir2017low-dvs-gesture}. 
\revcg{Here and in Figure~\ref{fig:impact_abo}, ``speed-up'' is defined as the ratio between the vanilla 8-bit bit-serial compute latency and the BitFair compute latency under the same layer schedule. It is computed from the number of bit-serial calculation cycles, which decreases as the early-termination rate increases.}
Conservative settings (e.g., $\theta = -10$) rarely trigger early termination, achieving 97.22\% accuracy but no speed-up (1.00$\times$). Relaxing the threshold increases early termination and improves speed-up to 2.12$\times$ at $\theta = \rev{-1.2}$ and 2.47$\times$ at $\theta = -0.5$, with accuracy degrading to 92.33\% and 90.65\%, respectively. Beyond this regime, accuracy drops sharply: at $\theta = 0.5$, accuracy falls to 70.03\% (2.79$\times$), and at $\theta = 2.5$ it collapses to 12.26\% even though speed-up saturates near 2.98$\times$. This sensitivity motivates BitFair's learnable, layer-wise thresholds, which avoid manual tuning while optimizing efficiency under an accuracy constraint.
\subsubsection{Impact of Adaptive Bit Ordering}
Figure~\ref{fig:impact_abo} quantifies the benefit of \ac{ABO}. Compared to fixed MSB-first ordering, \ac{ABO} consistently improves speed-up (e.g., from 1.68$\times$ to 1.74$\times$ on N-MNIST and from 1.83$\times$ to 1.92$\times$ on SVHN) while preserving or improving accuracy. The largest gain is on DVSGesture, where speed-up increases from 1.89$\times$ to 2.12$\times$ and accuracy improves from 95.91\% to 96.53\%. MNIST shows a similar trend (1.75$\times$ to 1.78$\times$, 96.88\% to 97.46\%), while N-MNIST also benefits in accuracy (97.03\% to 97.72\%). \rev{\ac{VWW} follows the same trend: speed-up increases from 1.65$\times$ to 1.80$\times$, and accuracy improves from 85.87\% to 86.52\%.} These results confirm that MSB-first ordering is not generally optimal, and that dataset- and layer-aware bit ordering can prioritize informative bits early to amplify early-termination opportunities without sacrificing accuracy. 


Notably, DVSGesture and N-MNIST are event-camera datasets that inherently exhibit high temporal sparsity, which naturally favors early termination. Therefore, we include SVHN, a dense RGB digit-recognition task where activations are substantially less sparse. Pure SNN solutions typically lose efficiency on dense inputs because high spike rates significantly increase synaptic event processing, reducing the sparsity advantage of event-driven computation. In contrast, BitFair still delivers 1.92$\times$ speed-up at 91.83\% accuracy on SVHN, demonstrating that its bit-serial early termination remains effective even when the input lacks event-driven sparsity.

\subsubsection{Comparison with CNN Early-Termination Baselines}
Table~\ref{tab:cnn_et_comparison} summarizes the CNN early-termination design space across all \rev{five} datasets. An 8-bit vanilla bit-serial scheme without early termination serves as the baseline. A BitSET~\cite{pan2023bitset}-style baseline, which employs early termination with MSB-first ordering and fixed thresholds derived from batch normalization, can improve speed-up but at the cost of reduced accuracy. BitFair consistently achieves \changed{the best} accuracy--speed-up trade-off by combining learnable thresholds with \ac{ABO}, reaching up to 2.12$\times$ on DVSGesture with only 0.58\% accuracy degradation from the vanilla \changed{bit-serial} baseline. \rev{Table~\ref{tab:training_cost} reports the offline training cost in wall-clock seconds on a single NVIDIA RTX 4090 GPU. BitFair's full pipeline, including learnable thresholds and the bit-ordering search, takes 2.4--4.5$\times$ the vanilla QAT time. This is consistently lower than the 3.6--6.1$\times$ overhead of the BitSET-style baseline, mainly because BitFair avoids BitSET's costly encoding-aware retraining step. As summarized by the $\Delta_{\text{vs.\,BitSET}}$ row in Table~\ref{tab:training_cost}, BitFair reduces the offline training time by 26.2--47.4\% relative to BitSET across the five datasets, for an average saving of 35.1\%. Note that this offline training cost is a one-time expense, which does not affect the inference speed-up.}

\begin{table}[t]
    \centering
    \caption{\rev{Offline training cost across datasets, measured in wall-clock seconds on a single NVIDIA RTX 4090 GPU.}}
    \label{tab:training_cost}
    \small
    \setlength{\tabcolsep}{4pt}
    \resizebox{\columnwidth}{!}{%
    \rev{%
    \begin{tabular}{@{}lccccc@{}}
    \toprule
    Method & MNIST & N-MNIST & DVSGesture & SVHN & \acs{VWW} \\
    \midrule
    Vanilla$^{1}$ & 324 & 1620 & 1675 & 5885 & 1092 \\
    BitSET~\cite{pan2023bitset} & 1168 & 9072 & 10215 & 33544 & 4696 \\
    \textbf{BitFair (Ours)} & 779 & 5994 & 7535 & 17654 & 3058 \\
    \midrule
    $\Delta_{\text{vs.\,BitSET}}$\,(\%)$^{2}$ & $-33.3$ & $-33.9$ & $-26.2$ & $-47.4$ & $-34.9$ \\
    \bottomrule
    \end{tabular}}}
    \par
    \vspace{2pt}
    \begin{minipage}{\columnwidth}
    \footnotesize
    \rev{\textsuperscript{1}Vanilla denotes an 8-bit quantization-aware training (QAT) without early termination.\\
    \textsuperscript{2}$\Delta_{\text{vs.\,BitSET}}=(T_{\text{BitFair}}-T_{\text{BitSET}})/T_{\text{BitSET}}\times100\%$ is the relative training-time change of BitFair over the BitSET-style baseline. Negative values indicate time saved (35.1\% on average).}
    \end{minipage}
\end{table}


\begin{figure}[t]
    \centering
    \includegraphics[width=0.99\linewidth]{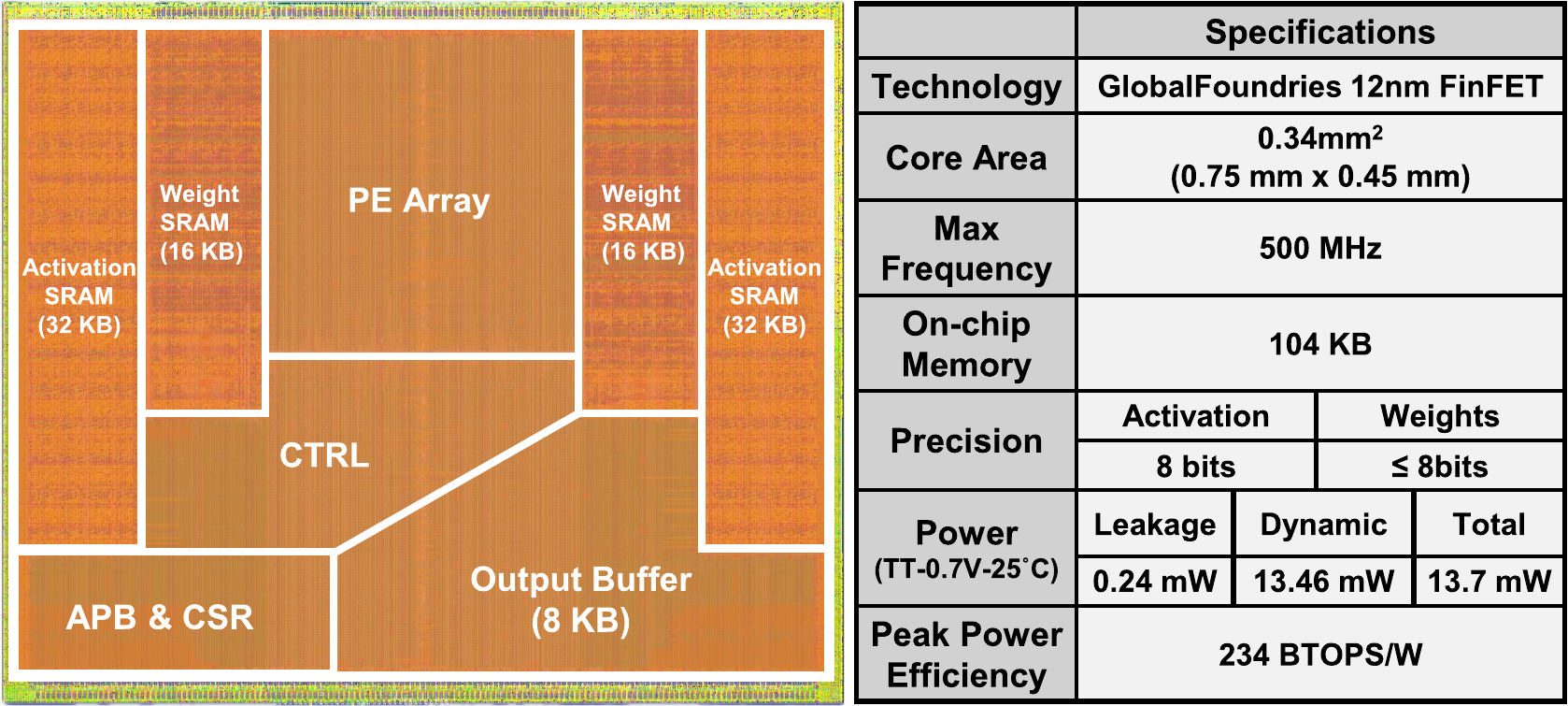}
    \caption{BitFair layout and post-layout specification.}
    \label{fig:chip_layout}
\end{figure}

\subsection{Hardware Implementation Results}
Figure~\ref{fig:chip_layout} shows the post-layout diagram and specification of BitFair. Implemented in 12-nm FinFET technology, BitFair occupies 0.34~mm$^2$ and reaches \changed{a maximum} frequency of 500~MHz. The 104~KB on-chip memory provides sufficient storage for the target edge vision tasks while maintaining a small silicon footprint. Post-layout simulation under a 0.70~V supply voltage at the typical process corner (TT) and 25$^\circ$C yields a power consumption of 13.7~mW at \revfinal{149.6}~BTOPS/W. Peak power efficiency of \revfinal{234.0}~BTOPS/W is achieved at the 0.55~V/200~MHz operating point (3.5~mW), where the BTOPS metric captures the bit-serial processing of quantized weights and activations~\cite{shi2025sparsecol}.

\subsubsection{Area breakdown.} The component-level area breakdown is shown in Table~\ref{tab:area_breakdown}. The total cell area is 0.277~mm$^2$, with the memory subsystem dominating at 51.2\%, followed by the PE array at 44.4\%. The controller and other peripheral logic together account for only 2.9\%, reflecting the lightweight control overhead of the bit-serial architecture. 

\subsubsection{Power Analysis.} 
\begin{figure}[t]
    \centering
    \includegraphics[width=0.66\linewidth]{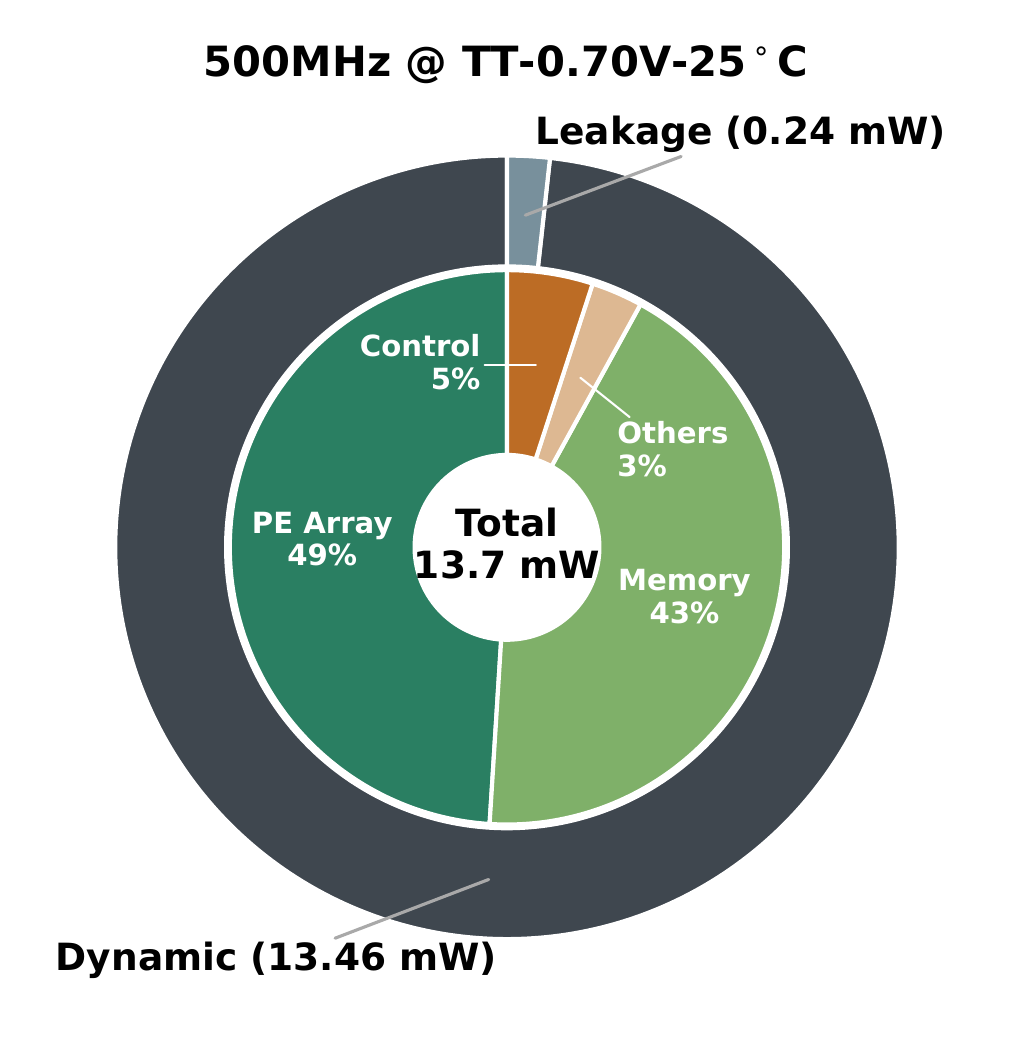}
    \caption{Power breakdown at 500\,MHz under the typical corner (TT, 0.70\,V, 25$^\circ$C). The outer ring distinguishes dynamic (13.46\,mW) from leakage (0.24\,mW) power; the inner ring decomposes dynamic power by functional block.}
    \label{fig:power_split}
\end{figure}

Figure~\ref{fig:power_split} provides a finer-grained power decomposition at the nominal 500\,MHz operating point under the typical corner (TT, 0.70\,V, 25$^\circ$C), where the total power consumption is 13.7\,mW. At this corner, dynamic power overwhelmingly dominates at 13.46\,mW (98.2\%), while leakage contributes only 0.24\,mW (1.8\%). The PE array accounts for the largest share (49\%), followed by the memory subsystem (43\%), control logic (5\%), and other peripheral circuitry (3\%).

\rev{The memory subsystem occupies 51.2\% of the cell area (Table~\ref{tab:area_breakdown}) and consumes 43\% of the dynamic power, about 5.8\,mW of the 13.46\,mW dynamic budget, as shown in Figure~\ref{fig:power_split}. Because dynamic power accounts for more than 98\% of the total power at this corner, this share mainly reflects active read/write energy rather than standby leakage. Several architectural choices keep this energy bounded. First, the 104\,KB on-chip SRAM stores target-network weights and activation buffers, eliminating off-chip DRAM traffic during inference for these workloads. Second, the output-stationary dataflow keeps each partial sum in a local accumulator, minimizing write-backs. Third, activations are broadcast across PE rows while weights are reused down columns, reducing redundant SRAM reads.}
\revcg{Early termination further lowers memory-related switching activity: once a PE terminates, the remaining weight-bit operands for that output are no longer consumed by that PE, and the local operand latches and datapath can be gated. When all PEs sharing a weight-memory access have terminated for the current bit plane, the corresponding SRAM read can also be suppressed. Otherwise, the read is amortized over the still-active PEs. Thus, the bit-cycle speed-ups reported in Table~\ref{tab:cnn_et_comparison} translate into reduced weight-operand delivery and datapath activity, with the exact SRAM-read reduction depending on tile-level termination patterns.}

\begin{figure}[t]
    \centering
    \includegraphics[width=\linewidth]{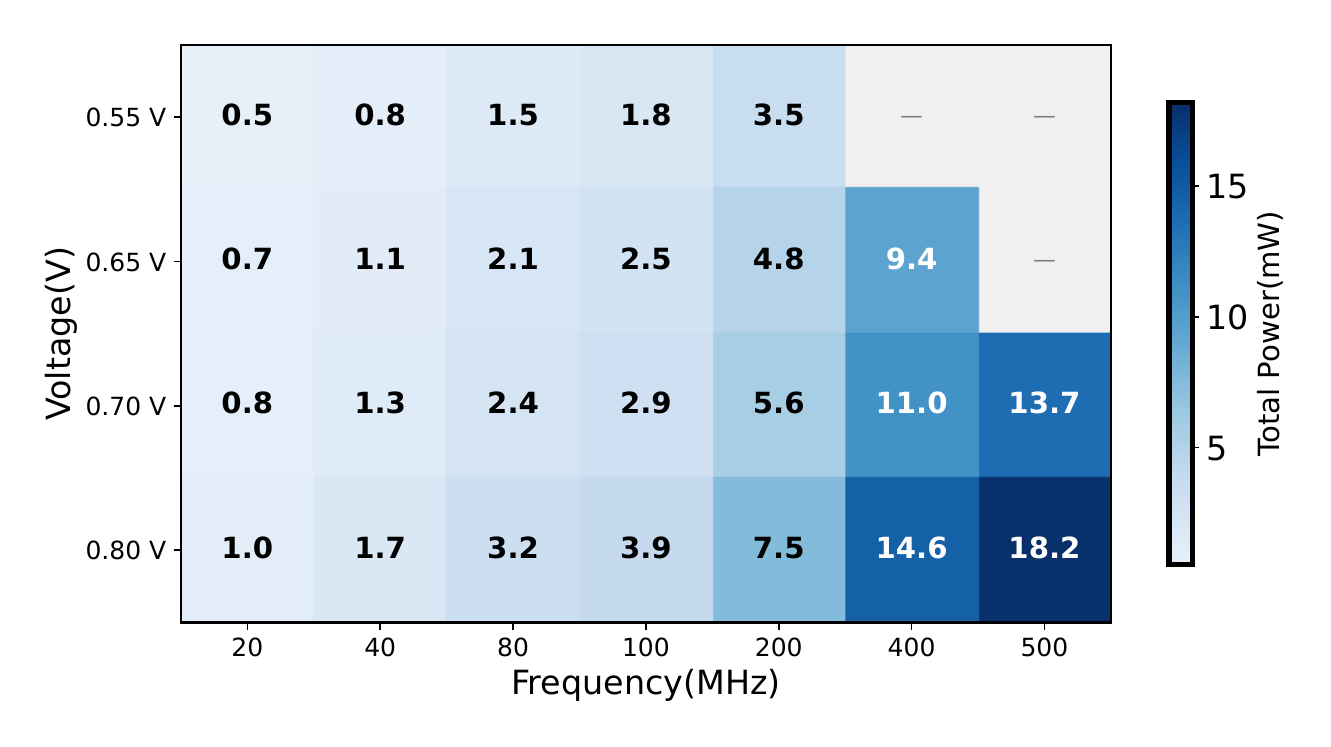}
    \caption{\changed{Power scaling heatmap across supply voltages and frequencies at TT corner.}}
    \label{fig:power_scaling_heatmap}
\end{figure}

To characterize the design's adaptability to varying power envelopes, \rev{we performed power analysis based on post-layout parasitic extraction} at multiple voltage and frequency points. As Figure~\ref{fig:power_scaling_heatmap} shows, BitFair exhibits a wide dynamic power range spanning from sub-milliwatt operation (0.5\,mW @20\,MHz, 0.55\,V) to 18.2\,mW at the highest operating point (500\,MHz, 0.80\,V). Reducing the supply voltage constrains the maximum achievable frequency yet substantially lowers total power and further enhances energy efficiency. This broad voltage-frequency scaling range enables dynamic voltage-frequency scaling strategies tailored to real-time XR workload demands: lighter tasks such as idle tracking can operate at reduced voltage and frequency to extend battery life, while computationally intensive gesture recognition bursts can leverage the full 0.80\,V, 500\,MHz operating point for maximum throughput.

\begin{table}[t]
    \centering
    \caption{Cell area breakdown of the BitFair accelerator.}
    \label{tab:area_breakdown}
    \small
    \begin{tabular}{lcc}
    \toprule
    \textbf{Component} & \textbf{Area (mm$^2$)} & \textbf{Percentage} \\
    \midrule
    PE Array   & 0.123 & 44.4\% \\
    Memory     & 0.142 & 51.2\% \\
    Control    & 0.008 &  2.9\% \\
    Other      & 0.004 &  1.5\% \\
    \midrule
    \textbf{Total} & \textbf{0.277} & \textbf{100\%} \\
    \bottomrule
    \end{tabular}
\end{table}

\subsection{Real-Time Latency Analysis for XR Interaction}
Extended reality applications, particularly natural hand-gesture \changed{recognition}, require sub-millisecond inference to preserve visual-motor synchrony and avoid simulator sickness. We evaluate BitFair on DVSGesture as a representative event-driven workload and report end-to-end throughput in Table~\ref{tab:sota_comparison}. Converting these throughput numbers to per-frame latency, BitFair achieves 645~FPS at 0.55~V/40~MHz (1.55~ms), 3225~FPS at 0.55~V/200~MHz (0.31~ms), and 8060~FPS at 0.70~V/500~MHz (0.12~ms). These results show that BitFair meets the stringent $<1$~ms XR latency target at moderate-to-high operating points, while still offering a low-power mode ($<1$~mW) when latency constraints are relaxed.

\subsection{Comparison with State-of-the-Art}
While several bit-serial architectures exist~\cite{sharify2019laconic, shi2024bitwave, wang2025bitpattern}, they target high-performance computing rather than energy-constrained edge applications. Furthermore, most report synthesis-based estimates without post-layout or silicon validation. Therefore, we compare BitFair against \changed{fabricated} state-of-the-art edge vision accelerators, as shown in Table~\ref{tab:sota_comparison}. These works primarily employ \ac{SNN} architectures targeting the same application domain. \changed{As analyzed in~\newref{\cite{ottati2023spike}}, the relative efficiency of \ac{ANN} and SNN accelerators depends on task type and data modality. Our results align with their finding that \ac{ANN}-based approaches can achieve superior efficiency on vision classification, while extending this to event-driven data with bit-serial early termination.}

As shown in Table~\ref{tab:sota_comparison}, BitFair achieves the highest accuracy on both datasets: 97.7\% on N-MNIST and 96.5\% on DVSGesture. At the 0.55~V operating point, BitFair achieves superior energy efficiency of 0.07~pJ/SOP, representing 4.0$\times$ and 8.9$\times$ improvements over~\cite{fu2025neuc} (0.28~pJ/SOP) and~\cite{yang202540nm} (0.62~pJ/SOP), respectively. Using the bit-level throughput power-efficiency metric adopted in~\cite{shi2025sparsecol}, BitFair achieves \revfinal{234.0}~BTOPS/W, the highest among all compared works, demonstrating 9.1$\times$ and 22.1$\times$ improvements over~\cite{yang202540nm} (\revfinal{25.8}~BTOPS/W) and~\cite{zhang202322} (\revfinal{10.6}~BTOPS/W), respectively. For the energy-delay product (EDP), BitFair achieves 0.21~nJ$\cdot$s on the DVSGesture~\cite{amir2017low-dvs-gesture} dataset at the 0.70~V operating point, representing 2.6$\times$ and 24.9$\times$ improvements over~\cite{yang202540nm} and~\cite{zhang202322}, respectively.

Implemented in 12-nm FinFET technology, BitFair achieves the smallest area \changed{of} 0.34~mm$^2$ and the highest frequency \changed{of} 500~MHz among the compared works. \changed{We evaluate performance across multiple voltages and frequencies to characterize power scaling.} The lowest voltage point yields the best energy efficiency, while the 0.70~V point maximizes throughput and \changed{minimizes} EDP. \changed{Although} the absolute power at higher voltages exceeds that of ultra-low-power designs~\cite{frenkel2022reckon}, this is offset by higher throughput and lower latency. When normalized by computational throughput, BitFair demonstrates superior efficiency across all operating points.

\section{\rev{Limitations and Future Work}}
\rev{BitFair's efficiency gains come from ReLU-induced dynamic sparsity that early termination can skip later bit-cycles only for outputs that ReLU will eventually clamp to zero. The attainable speed-up is therefore bounded by the fraction of negative pre-activations in each layer and is highest when this sparsity is high. When activation sparsity is low, whether because of the layer type, model architecture, or dataset, there are fewer opportunities to terminate early. In these cases, speed-up degrades gracefully toward the vanilla bit-serial baseline ($1.00\times$), while accuracy is preserved because the learned thresholds suppress only reliably negative outputs. This trend appears in our results. A high-sparsity event-camera workload such as DVSGesture yields the largest gain, with a $2.12\times$ speed-up and an average model ReLU activation sparsity of $68\%$. The denser RGB SVHN task, with a lower model ReLU activation sparsity of $55\%$, shows a more modest but still useful $1.92\times$ speed-up while maintaining $91.83\%$ accuracy. The mechanism also relies on activation functions with a hard-zero region. Networks using smooth non-ReLU activations, such as GELU, rarely produce exact zeros and would benefit less.

Nevertheless, the core ideas of early termination and adaptive bit ordering remain applicable beyond the present network size, which is tailored to lightweight visual perception under tight on-chip memory and input-resolution constraints. In the future, we plan to scale it to heavier XR workloads, such as pose estimation or object detection on the COCO dataset~\cite{lin2014microsoft}. Other potential directions include combining BitFair's dynamic early termination with static bit-level sparsity and adapting it for non-ReLU activations.}

\section{Conclusion}
\label{sec:conclusion}
This paper presents BitFair, an \changed{ultra-low-power} bit-serial \changed{CNN} accelerator that balances accuracy and efficiency by employing learnable per-layer thresholds and adaptive bit ordering. BitFair \changed{uses} gradient-based optimization to learn layer-specific thresholds and employs a greedy search to identify bit orders that maximize early termination. \changed{Implemented in} GlobalFoundries 12-nm FinFET technology, BitFair demonstrates 4.0$\times$ to 22.1$\times$ energy efficiency improvements over state-of-the-art edge vision accelerators, with an accuracy loss of \changed{less than} 1.6\% compared to floating-point baselines, validating our approach for \changed{resource-constrained XR vision applications}.

\section{Acknowledgment}
\revfinal{This work was partially supported by the Dutch Research Council (NWO) under the Talent Programme Veni 2023 scheme in Applied and Engineering Sciences (AES), Grant No. 21132 (Energy-Efficient Real-Time Edge Intelligence for Wearable Healthcare Devices).} We thank GlobalFoundries for providing us the 12LP-PLUS PDK through the GF12+ University Partnership Programme.

\bibliographystyle{IEEEtran}
\bibliography{references}

\vspace{-30pt}
\begin{IEEEbiography}[{\includegraphics[width=1in,height=1.25in,clip,keepaspectratio]{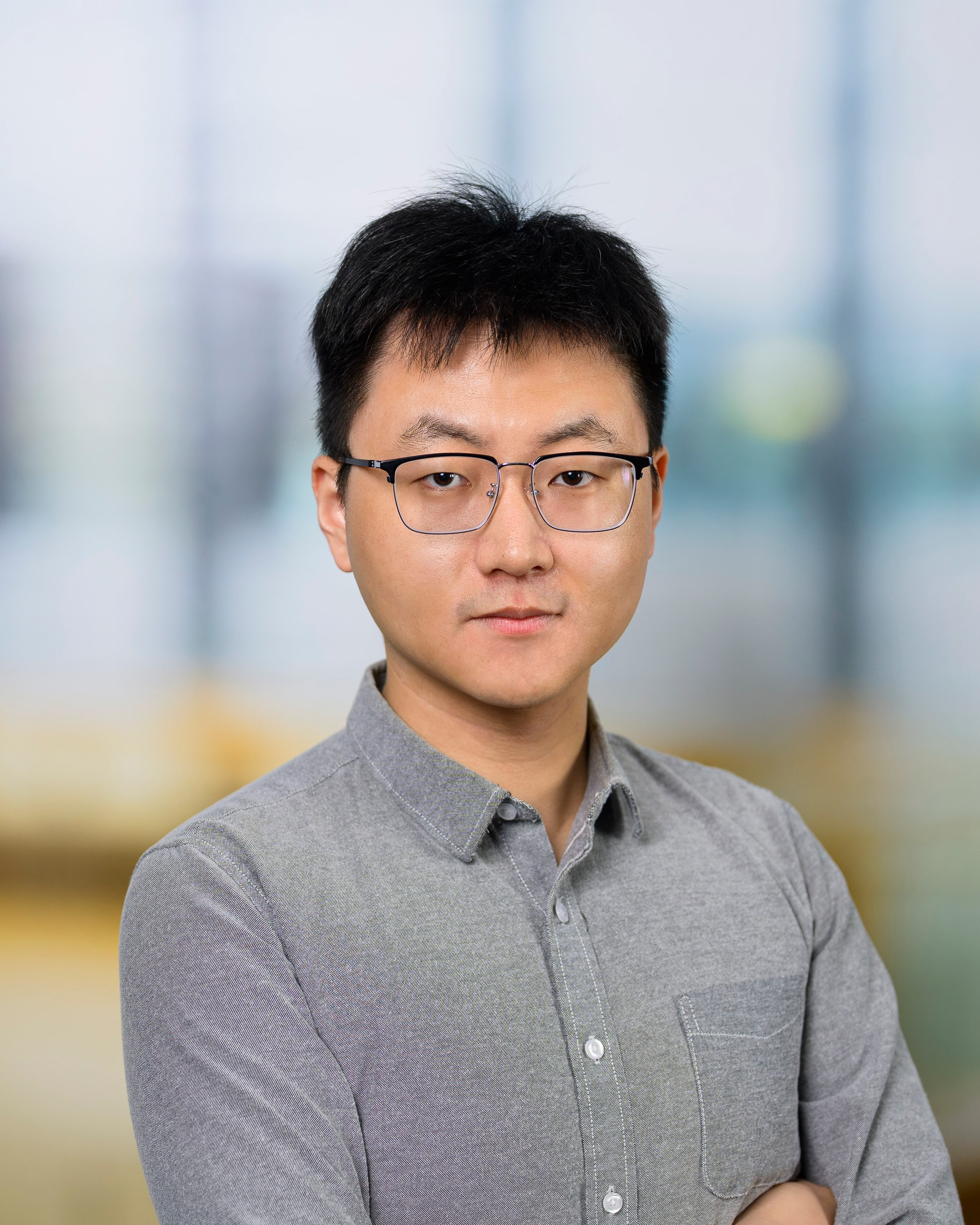}}]{Ang Li} (Student Member, IEEE) received the B.S. degree in Microelectronics from the School of Microelectronics, Xidian University, Xi’an, China, in 2019, and the M.S. degree in Integrated Circuits from the School of Integrated Circuits, Tsinghua University, Beijing, China, in 2022. He is currently pursuing a Ph.D. in Microelectronics at Delft University of Technology, The Netherlands. His research interests include deep learning, digital circuit design, computer vision, and VLSI digital signal processing.

\end{IEEEbiography}

\begin{IEEEbiography}[{\includegraphics[width=1in,height=1.25in,clip,keepaspectratio]{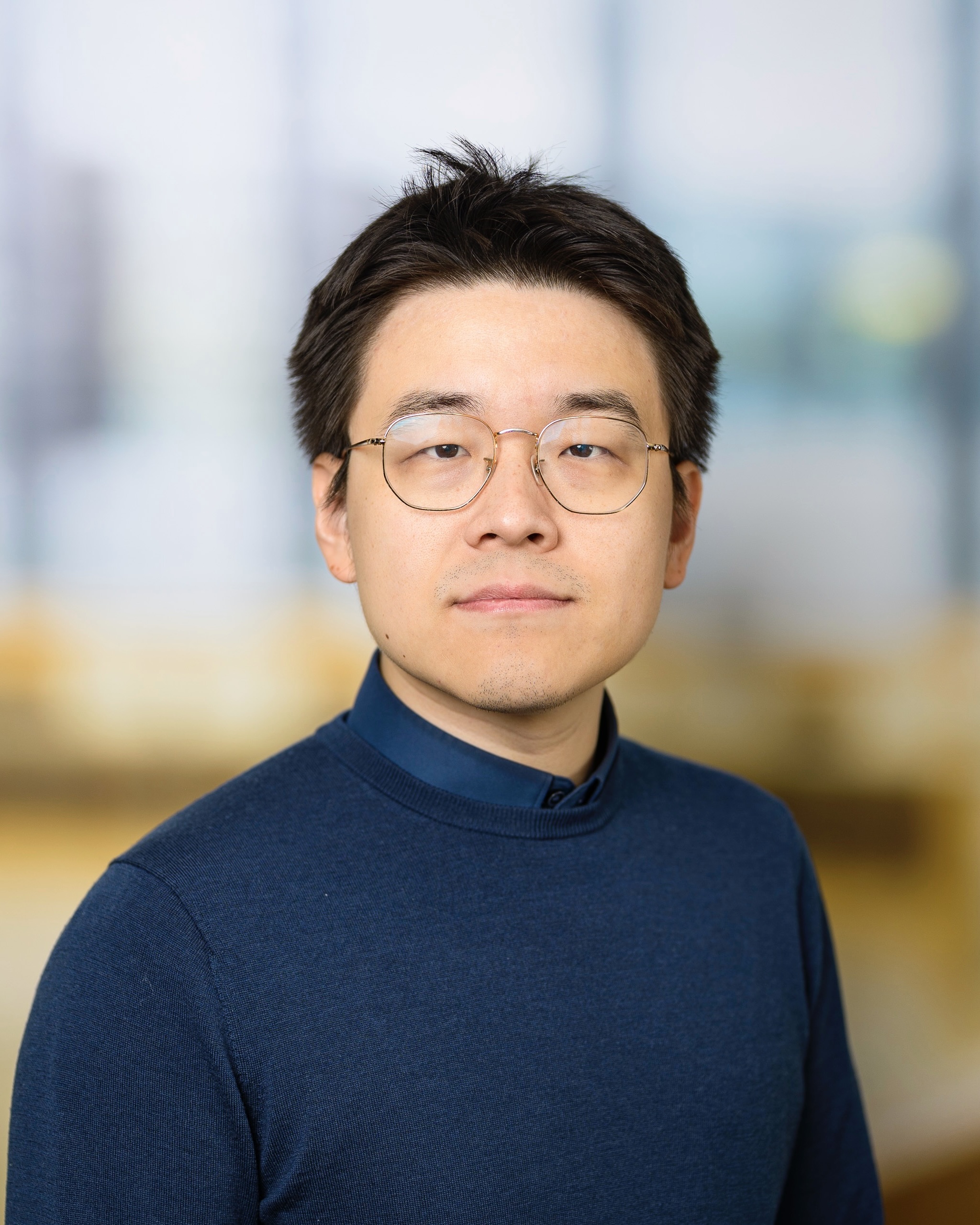}}]{Chang Gao} (Member, IEEE) received his Ph.D. degree with distinction in Neuroscience from the Institute of Neuroinformatics, University of Zürich and ETH Zürich, Zürich, Switzerland, in March 2022. He received his M.Sc. degree from Imperial College London in September 2016 and his B.Eng. degree from the University of Liverpool and Xi’an Jiaotong–Liverpool University in July 2015. In August 2022, he joined Delft University of Technology, The Netherlands, as a tenured Assistant Professor in the Department of Microelectronics. He leads the Lab of Efficient Machine Intelligence (EMI), where he conducts research on hardware–software co-design for edge AI computing and embodied intelligence. He received the 2022 Misha Mahowald Early Career Award in Neuromorphic Engineering and a 2022 Marie Skłodowska-Curie Postdoctoral Fellowship. He is a 2023 Dutch Research Council (NWO) Veni laureate and a 2023 MIT Technology Review Innovator Under 35 in Europe for his contributions to algorithm–hardware co-design for efficient sparse recurrent neural-network edge computing.
\end{IEEEbiography}

\end{document}